# Gate Design and Stage-Dependent Incentives in Retail Proprietary-Trading Evaluations

*Why passing is not standalone evidence of skill, and why the product fails to pay under measured trading constraints*

Nicholas Hall

## Abstract

Retail proprietary-trading firms sell a two-stage product: a paid evaluation that must reach a profit target before breaching a trailing drawdown, followed by a funded account that must survive a minimum window and satisfy a consistency rule before a payout is released. We show that the geometry of this contract creates incentives that differ by stage and that make passing a poor standalone signal of skill. Under end-of-day trailing the evaluation rewards a fast, lumpy trade cadence while the funded account punishes it, by a factor of nine in the joint gate at identical strategy parameters. The evaluation is defeatable at zero skill: position sizing alone yields a pass probability near 0.40, against a measured cohort rate of 0.168. Pass probability does rise with skill, but at a fixed cadence a real edge and aggressive sizing move it by nearly the same amount, so a pass rate confounds the two and cannot by itself be read as evidence of skill. And under a simplified contract model the seller's contribution margin is bounded by the gap between perceived and actual gate probabilities — the probability analogue of shrouding a price component — under which a permeable, marketed evaluation and a hard, unpriceable payout gate are what a margin-maximising seller would choose. The observed design is consistent with that: across the sector, pass rates are published far more often than payout rates, and two families of firms with inverted rule architectures differ by a factor of 1.7 in pass rate within a single sample, an association consistent with substitution between gates rather than with skill filtering.

The same geometry produces negative expected value for the participant. Within the strategy universe we measured, and under the trade-level cost model the paper relies on, no measured strategy configuration at the drifts we observed clears break-even on any account structure we could source: a transaction-cost floor exceeded every gross per-trade edge surviving our pre-specified research protocol, and the joint pass-and-payout gate is compressed to at or below the line implied by the account price. Under the baseline simulation model, break-even lies between a 40.5% and 41.5% win rate at 1:1.5 net of costs across the three accounts studied in detail and the two implementations, against a driftless baseline of 40.0%. We also measure the delta-neutral construction these firms prohibit and find it approximately expected-value neutral under symmetric sizing at the observed payout ceiling, with the ceiling rather than the prohibition doing the work. Every account-level result is reported against a zero-edge control through the identical pipeline, without which pass and payout rates are uninterpretable.

## 1. Introduction

A retail proprietary-trading evaluation is a product, not an employment screen, and the distinction matters for how it should be analysed. The participant pays a fee in the range of roughly \$87 to \$228 for an account with a simulated balance, a profit target, and a maximum-loss level that trails the account's equity peak. Reaching the target converts the account to "funded"; breaching the loss level terminates it. A funded account is subject to a further set of gates — a minimum number of trading days, a minimum number of winning days, and in many cases a consistency rule capping the fraction of total profit any single day may contribute — before a payout is released.

The industry's public framing is that these constraints screen for discipline. This paper's central finding is that they do something else: they create incentives that differ between the two stages and that make the act of passing a poor guide to the possession of skill. The constraints function as the product's pricing mechanism, and the resulting structure is one in which the stage the participant can see is permeable and the stage that pays is not. The firm sells a path-dependent, capped-payoff contract with option-like characteristics: downside to the buyer capped at the fee, upside capped at the payout ceiling, and the gate parameters setting the effective strike. Under that reading, the question "is the evaluation passable?" is the wrong question, because it is trivially yes, and the right question is what the geometry rewards at each stage and whether the joint probability of passing and then extracting exceeds the ratio of cost to payout size. Both are quantities this paper measures.

### 1.1 The two binding constraints

The paper's conclusion rests on two constraints that are independent of one another. Each independently constrains feasibility; together they produce the result.

- The cost floor. Measured round-turn friction on the micro contract is approximately 1.12 points, or $2.24, comprising commission and measured slippage. The largest gross per-trade edge measured in this programme at bar level, across every candidate surviving the pre-specified protocol of §2.3, is +0.32 points. On the full-size contract the friction is $13.80 per round turn at the base estimate and $28.80 at the stressed ninetieth percentile. No measured candidate is net-positive after costs.
- Gate compression. The joint hit rate — pass probability multiplied by payout probability conditional on passing — is set by the account geometry to a level at or below the break-even line implied by the account price, at the measured participant drifts, on every firm modelled. Under the seller model of §13, a gate clearing comfortably above the seller’s own break-even would be inconsistent with a non-negative contribution margin; Table 12 is the empirical evidence that it does not.

The two constraints have no shared parameter. The cost floor is a property of the exchange and the broker; the gate is a property of the contract the firm writes. A participant improving one does not relax the other. The first is an empirical statement about the strategies we measured and cannot be more than that; the second is a property of the contract — the geometry is fixed by the seller — although whether it compresses expected value below break-even depends on the participant’s drift, as §11 makes explicit. A reader who rejects every strategy in §8 as poorly chosen still faces §5 and §7.

### 1.2 Contributions

Listed with the results on gate design first, since they hold for any strategy, and the results on expected value second, since they are conditional on the strategies measured.

1. A frequency asymmetry result: the evaluation stage and the funded stage require opposite trade cadences, by a factor of nine in the joint gate at identical win rate. This mechanically explains a widely-practised but unexplained industry convention (§6).
2. A demonstration that evaluation pass rates are manufacturable by position sizing alone: the rules permit a zero-skill participant with free sizing a pass rate of roughly 0.36 to 0.47 under end-of-day monitoring, against an observed cohort rate of 0.168, which implies the evaluation pass rate cannot by itself serve as a reliable skill filter and any consequential filtering not explained by it must occur at the payout gate (§5.3).
3. An external check of the funded-stage payout rate against a futures firm’s published cohort statistics on identical rule geometry, agreeing to within roughly 1.5 percentage points, and the observation that two families of firms with inverted rule architectures differ materially in pass rate, while separate sector datasets show downstream payout incidence remains low — evidence consistent with substitution between gates rather than with skill filtering (§9.1–9.3).

4. A methodological argument that evaluation-account results are uninterpretable without a zero-edge control through the identical pipeline, since the reported quantity is the sum of the account's option value and the strategy's contribution, and the two are not separable from the total (§2.4).
5. A simulation study over the (μ, σ, f) surface that replaces the single worked example of earlier drafts and generalises the feasibility statement from an anecdote to a property of the surface (§5).
6. The cost–cadence pincer: a characterisation of the infeasible region as the intersection of a cost floor pushing required gross edge up and a cadence requirement pushing required pass probability and speed up (§5.5).
7. Fee sensitivity: the demonstration that the account fee moves the break-even threshold by two to three points of win rate across the observed price range but, at the drifts anyone has demonstrated, does not overturn the negative sign, because the fee does not enter the drift (§11).
8. An adversarial empirical test in which the most prominent public practitioner's fully-specified rule set, implemented without interpretation, is shown to be coin-equivalent (§8.5).
9. A measurement of the hedged-pair construction — the delta-neutral attack the firms prohibit — showing it to be expected-value neutral at the observed payout ceiling, and identifying the ceiling rather than the prohibition as the binding constraint (§12).

### 1.3 Related literature

**We are not aware of a refereed treatment of this product.** That individual traders lose money in aggregate is among the better-established findings in empirical finance — Barber and Odean (2000) and Odean (1999) on overtrading, and Barber, Lee, Liu and Odean (2009) on the aggregate magnitude of retail losses. What has not been studied is the product examined here. Retail proprietary evaluations have been the subject of industry reporting, practitioner commentary and regulatory consultation, but we could locate no peer-reviewed study of their expected value, their rule geometry, or the economics of the firms offering them. The absence is notable given the sector's size and the regulatory attention documented in §9.6. This paper is therefore positioned against three adjacent literatures rather than a direct one.

The closest ancestor is the literature on shrouded attributes. Gabaix and Laibson (2006) show that a firm selling a base good with an associated add-on will, in equilibrium, price the visible component low and the shrouded component high, and that this survives competition, costless advertising, and the presence of sophisticated consumers who understand the scheme. Our Proposition 5 is the probability analogue of that result rather than an independent one. What is shrouded here is not a second price but the likelihood of receiving the promised payoff: the evaluation fee is the visible base price, and the payout gate — a compound of a safety net, a qualifying-day count, a consistency test and a ceiling — is the attribute the buyer cannot price at the point of purchase. The equilibrium shape is the same and we claim no priority for the mechanism. The contribution is that the shrouded attribute is a probability, that it is recoverable from published rules, and that we recover it.

The second is the pricing of fixed-odds betting products. Makropoulou and Markellos (2011) model optimal price setting by a bookmaker facing information uncertainty, which is structurally the problem of §13.1: a seller choosing terms against a population whose beliefs about the odds differ from the odds. Constantinou and Fenton (2013) document exploitable biases in the same markets, the counterpart to the hedged construction of §12. The analogy is close enough to state plainly — the product studied here is a fixed-odds proposition whose observed performance measure conflates skill with position sizing, and the 0.71% of funded participants reaching real capital reported in §9.1 is difficult to reconcile with any other reading.

The third is drawdown-constrained betting, where the classical reference remains Markowitz (1952) on the utility of wealth, with Bailey and López de Prado (2013) the closest modern treatment of stop-outs defined on drawdown rather than on level. The first-passage result of §4.2 is standard there and is stated, not claimed. Our contribution in that section is the measurement of the ratchet penalty — the fourteen to seventeen points of pass probability consumed by a barrier that trails the running peak rather than sitting at a fixed level — which is a property of the specific contract rather than of the mathematics.

**Against that background, what is new here is empirical rather than theoretical.** Three claims we believe are not in the literature: that evaluation architecture is strongly associated with the pass rate — documented in §9.3 from a within-population split reporting 20.3% for one-phase against 11.8% for two-phase evaluations; that the disclosure pattern across the sector — pass rates published, payout rates much less frequently disclosed — matches what the shrouding account predicts, documented in §9.2; and that the evaluation gate is defeatable at zero skill by position sizing alone, with the measured cohort performing worse than that construction, which §5.3 derives and §9.1 corroborates.

> *Citation discipline. The references above were verified individually against the published record. No reference in this paper is generated, inferred from a secondary mention, or reconstructed from memory. Appendix B.1 records citation status: every claim relied upon in this draft has a verified source, and the two that earlier drafts carried on self-report were removed rather than sourced retrospectively.*

## 2. Apparatus and search protocol

A null result is only as credible as the search that produced it, and the objection that the search was too shallow cannot be answered by adding results — only by documenting the protocol under which they were generated and discarded. This section describes the apparatus, the gates every candidate passed, and the control against which every candidate was measured.

### 2.1 What the apparatus was built for

It was not built to write this paper, and we state that plainly because the alternative framing would be a retrofit. The apparatus was constructed to find a deployable trading edge. It ran three research cycles — bar-level strategies, evaluation-account extraction, and order-flow microstructure — and returned no deployable evaluation-compatible edge on any of them. This paper is what the null result became once it was clear the null was the finding.

The distinction matters for weighing the evidence. Infrastructure purpose-built to demonstrate that nothing exists is weak evidence, its incentives pointing at the conclusion. Infrastructure built at cost to find something, run under pre-specified kill criteria, and returning nothing is considerably stronger. It is the second.

### 2.2 Structure

Two layers. Beneath, a strategy-agnostic optimisation harness performing parameter sweeps under a fixed in-sample/out-of-sample split, selecting in sample only and reporting out-of-sample survival. Above it, a multi-agent orchestration layer dispatching investigations and maintaining an auditable trail from every reported number to the run that produced it.

Eight operational agents ran in a defined reporting structure — research, backtesting, implementation, risk, and quantitative research functions, together with summarisation and reflection roles — all reporting to a coordinating agent, which in turn reported to the human principal. Decision authority was never delegated: kills, escalations, and any advancement toward execution required explicit human approval, and the risk function held veto power over the others.

| Quantity | Count |
|---|---|
| Tracked investigations | 127 |
| Resolved to a verdict | 88 |
| Research documents produced (with revisions) | 170 (416) |
| Full agent run transcripts retained | 349 |
| Attributed comments on investigations | 444 |

| Quantity | Count |
| --- | --- |
| Activity-log events | 1,928 |
| Operational agents | 8 |
| Model tokens consumed (input / output) | 814,037 / 6,816,502 |
| Active period | 2 – 21 July 2026 |

*Table 1. Apparatus throughput over the orchestrated phase. Every reported figure in this paper traces to a retained transcript. Source: export manifest, 2026-07-21.*

Two figures are worth drawing out. The entire program consumed approximately $113 of market data against a $250 cap, so the null result is not an artefact of an underfunded data budget — the data was never the binding constraint. And execution remained locked throughout: across 127 investigations no order was ever placed, live or simulated. Nothing in this paper is contaminated by a realised trading record, favourable or otherwise.

### 2.3 The gates

Every candidate passed through the same gates in the same order, fixed before any candidate was tested:

1. Mechanism. A candidate required a named structural counterparty — who is compelled to transact unfavourably, why the compulsion is durable, and in which regime it appears and disappears — before any code was written. Textbook patterns with no named mechanism were rejected without testing.
2. Maximum drawdown against the account limit. Drawdown is checked first because it is terminal: a candidate that breaches the limit is dead regardless of its returns.
3. Net edge per trade against the measured cost floor.
4. Out-of-sample Sharpe ratio against a floor of 1.5. The floor is a research decision rule fixed in advance, not a statistical significance threshold; Sharpe ratios are computed on per-trade returns and annualised by trade count, without a deflation for the number of trials, which §2.4 addresses by control rather than by correction.
5. Out-of-sample trade count, minimum thirty, before any candidate could advance.

The protocol follows the discipline urged by Bailey, Borwein, López de Prado and Zhu (2014), who show that an unrecorded number of backtest trials renders a reported Sharpe ratio uninterpretable, and whose probability-of-backtest-overfitting framework (2016) formalises the point. Selection and tuning occurred on in-sample data only. Each surviving candidate received exactly one out-of-sample validation for selection purposes, with no re-selection afterwards; a post-validation consistency audit, which can only kill, is permitted and was applied once (§8.1). The split date was fixed a priori at 7 September 2021 and never moved. Costs were applied from the first run of every sweep, and cost-free results were treated as void — a policy adopted after an early sweep of 576 opening-range configurations produced 190 survivors without costs and zero survivors with them.

**Kill criteria were pre-specified and kills were final.** A candidate that failed a gate was not re-parameterised and resubmitted, and closed questions were not reopened on the strength of a later idea. This is a constraint on the researcher rather than on the research, and it is the reason the 88 resolved investigations can be read as 88 separately resolved tests under a fixed protocol rather than as one search with many restarts. They are not statistically independent — they share data, code, market regimes and a research team — and no multiple-testing correction is claimed for them.

### 2.4 The zero-edge control

Every account-level result in this paper is reported alongside a control: a fair coin flip of fixed magnitude and random direction, executed through the identical pipeline, with identical costs, identical account rules, and identical gates. The control is not a robustness check appended at the end. It is the measurement.

The reason is the structural finding of Section 5.3. Because a capped-downside evaluation account carries option value, and because pass rates are manufacturable by position sizing alone, a strategy can post a respectable-looking pass rate and payout rate while contributing nothing whatsoever. Absolute figures cannot distinguish the two cases. Only the difference against a driftless walk through the same machinery can. This is what converted an apparently encouraging 16.16% payout rate into the finding that the coin posts 12.50% and the strategy sits 0.10 percentage points below the coin on the joint gate.

We note that the retail backtesting literature, and the practitioner material this paper examines adversarially, generally reports no null model at all. Where a control is absent, a reported pass rate is uninterpretable, because the quantity being reported is the sum of the account's option value and whatever the strategy contributes, and the two are not separable from the total alone.

### 2.5 What ran inside the apparatus, and what did not

Two qualifications are required, and blurring either would misrepresent the record.

First, the orchestration layer produced the investigation ledger, but the two deciding measurements reported in Section 8 were executed directly against standalone scripts rather than dispatched through it. The payout-rate verdict reused core routines already ratified within the orchestrated phase; the adversarial test of Section 8.5 was written afterwards as a fresh implementation. That the adversarial test reproduces the earlier verdict from an independent codebase is a strength, but it is a strength of replication, not of orchestration, and should be read as such.

Second, the orchestration layer is not operational at the time of writing, having failed during session initialisation after the research phase concluded. The complete record — every investigation, document, transcript, and comment — was exported and preserved beforehand, and all figures in this paper derive from that export. No finding depends on the layer being runnable, but a reader should not infer a live system from this description.

### 2.6 Data and cost model

Continuous front-month index futures, one-minute open-high-low-close-volume bars, June 2010 to June 2026, approximately 5.3 million bars, with daylight-saving-correct conversion from UTC to Eastern Time. The in-sample and out-of-sample split at 7 September 2021 yields 2,894 in-sample and 1,241 out-of-sample sessions in the full ledger; the adversarial run of Section 8.5 used 2,832 and 1,217 after its own filtering. The backtest engine was calibrated against an independent commercial platform, agreeing to a profit factor of 0.970 against 0.957 over approximately 3,800 trades with identical trade endpoints.

Micro contract: point value $2.00. The base execution model used in the early sweeps carried $1.00 commission and $2.00 slippage per round turn, about 1.5 points or $3.00; the finalised model used for the payout-rate work, calibrated on the measured execution sample, carries $2.24. Both are reported where used. Full-size contract: $20 per point and $13.80 of measured round-turn friction at the base estimate, comprising commission and slippage, with a stressed ninetieth-percentile figure of $28.80.

One property of the cost structure drives much of the paper and should be stated separately, because it is instrument-specific and easily missed: **fixed commission is roughly ten times heavier in proportional terms on the micro contract than on the full-size contract.** High-frequency thin-edge strategies are therefore structurally dead on the micro contract before any question of skill arises, which is precisely the instrument a retail evaluation account pushes participants toward. A viable candidate must be either highly selective or must capture large moves; the middle of the distribution is unreachable.

Simulation parameters. Phase diagram: 4,000 Monte Carlo paths per grid cell, 2,500 in the sizing search, 90-day horizon, fixed seeds. Adversarial two-stage test: absorbing-barrier Monte Carlo, 40,000 paths per cell, 120-day horizon, fixed seed, runtime 74 seconds. Payout-rate optimisation: 700 in-sample and 1,500 out-of-sample bootstrap attempts per cell, grid fixed a priori.

## 3. The product: cost structure and rule surface

This section describes the object under study in enough detail to make the later results legible, because the argument turns on contract mechanics rather than on market behaviour. Every firm parameter below was compiled from live firm sites on 10 July 2026 and recorded with provenance. No firm parameter in this paper is assumed where a measured value exists, and where a rule could not be sourced it was flagged rather than inferred — a discipline adopted after early rule-surface modelling errors, all of which arose from assuming a parameter that had already been measured.

We study three firms in detail — Lucid Trading, Tradeify, and TopOne Futures — chosen to span the rule space rather than as a convenience sample. Between them they cover both account structures, the full range of consistency regimes from none to the tightest observed, accounts with and without daily loss limits, and a price range of roughly six to one. A fourth, Apex Trader Funding, is added in §12.6 as the limiting case on price. Firms are named throughout because every parameter reported here is published by the firm and a reader must be able to check it. Pricing is promotional and changes frequently, so all figures carry a retrieval date.

> *Scope of attribution. Named firms appear in this paper only as the source of published parameters and published statistics. Where later sections characterise the product — in particular the seller's problem in §13 — the analysis concerns the contract form that this sector has converged on, not the conduct, intent or good faith of any firm named here. We do not allege fraud, misrepresentation, or deliberate design against the participant. A contract can produce a negative expectation without anyone having set out to construct one, and nothing in our data distinguishes the two cases.*

### 3.1 Two stages, two games

The evaluation and the funded account are separate problems with separate optimal policies, and conflating them was a defect in earlier internal work that materially distorted the results. The evaluation stage is optimised to pass: reach the target before the maximum-loss level, with downside capped at the fee, so aggression is cheap. The funded stage is optimised to extract: survive the payout window, satisfy the consistency rule, and reach the withdrawal threshold. The two objectives point in opposite directions, and §6 quantifies by how much.

A participant may also skip the first stage entirely. Instant-funding accounts place the participant directly into a funded account at three to four times the price — approximately $364 to $407 at the 50K tier against $87 to $98 for an evaluation. The choice is not obviously resolved in either direction, because the evaluation route runs two gauntlets at low cost per attempt while the instant route runs one at high cost. Which dominates is an empirical question about cost per funded account, and it must be computed per firm rather than assumed.

### 3.2 The drawdown is end-of-day trailing

On all three firms the maximum-loss level is set from the highest end-of-day balance and updates only at session close. Intraday equity may fall below the prior peak provided it remains above (peak_EOD − d). This matters more than it appears. Under end-of-day trailing the loss level is static within a session, so additional intraday trades do not compound drawdown risk within that session; this is the mechanism behind the frequency asymmetry of §6. Earlier internal modelling used intraday trailing and produced a materially more pessimistic picture, including a blow-up rate near 72% that was an artefact of the wrong barrier. Correcting it moved the break-even threshold down by several points of win rate.

> *Scope note: firms using intraday trailing on unrealised profit and loss are strictly harsher than the class modelled here. Every no-go statement in this paper is therefore conservative with respect to that class.*

### 3.3 The barrier geometry is not constant across account sizes

Each account specifies a profit target T and a maximum drawdown d. Under the driftless limit of §4.2, the pass probability of a fixed-barrier account is d/(d+T), which depends only on the ratio of the two. That ratio is not held constant as account size increases.

| Account size | Profit target T | Max drawdown d | d/(d+T) |
|---|---|---|---|
| 25K | $1,500 | $1,000 | 0.400 |
| 50K | $3,000 | $2,000 | 0.400 |
| 100K | $6,000 | $3,000 – $3,500 | 0.333 – 0.368 |
| 150K | $9,000 | $4,000 – $5,000 | 0.308 – 0.357 |

*Table 2. Barrier geometry by account size, ranged across the three firms. The driftless pass probability falls monotonically as account size rises. Source: firm rules reference, 2026-07-10.*

**The smallest accounts carry the most favourable geometry, and it degrades monotonically with size.** A zero-skill participant on a 50K account faces a driftless pass probability of 0.400; the same participant on a 150K account faces as little as 0.308. The larger account offers a larger nominal payout and a strictly worse chance of reaching it. This is the first indication that the account parameters are set by the seller rather than derived from any risk-management rationale, and it is the reason the entire research program studied the 50K tier: it is the most favourable case among the three firms studied in detail, so a no-go established there is conservative across that set; §12.6 later tests a cheaper and less restricted account separately.

### 3.4 Consistency rules

A consistency rule caps the fraction of cumulative profit that any single day may contribute. Across the observed accounts the rule appears in three regimes: absent, a 40% cap, and a 20% cap. Its placement varies — some accounts impose it only in the funded stage, some in both, and one imposes none at all at the tiers studied. The property is fixed at purchase and cannot be negotiated.

Its cost is strongly non-linear in trade cadence, which is what makes it analytically interesting rather than merely restrictive. At one trade per day a 40% cap costs nothing, because a single daily trade cannot dominate cumulative profit across the required number of days. A 20% cap, by contrast, roughly halves the funded-stage payout rate at the lumpy sizing used in Table 3.

| Win rate | No consistency | 40% cap | 20% cap |
|---|---|---|---|
| 40.0% | 34.8% | 34.8% | 17.4% |
| 42.0% | 40.9% | 40.9% | 22.5% |
| 44.0% | 46.8% | 46.8% | 28.6% |
| 46.0% | 52.6% | 52.6% | 34.8% |

*Table 3. Funded-stage payout rate by consistency regime, at a sizing that produces lumpy daily profits. The cost of the rule is strongly sizing-dependent; see the discussion below. Source: two-stage simulation, this paper.*

The rule interacts with the profit distribution rather than with its mean. A configuration producing lumpy wins survives an unconstrained account and fails a 20% account at the same win rate; a steadier configuration clears the gate and extracts more slowly. Since the two objectives cannot be maximised together, the consistency regime effectively selects the strategy class before any consideration of edge.

**Whether the rule is worth paying to avoid depends on the configuration, and this is easy to get wrong.** The cheapest 50K evaluation observed is $87 and carries a 20% funded consistency rule; the next cheapest is $98 and carries none. Table 3 makes the rule look decisive, halving the payout rate. But Table 3 is computed at a sizing that produces lumpy daily profits. Re-solving the two-stage model at a steady configuration — one trade per day at moderate risk, accumulating the withdrawal threshold over eight to ten winning days — places the largest single day at roughly 18% of cumulative profit, just inside the 20% cap. Under that configuration the rule barely binds and the cheaper account dominates on break-even win rate, 40.5% against 40.9%.

**The correct statement is therefore conditional, and it is sharper than the unconditional one.** A consistency rule is not a fixed tax on an account; it is a constraint on the joint distribution of daily profits, and its cost is a function of how lumpy the participant's equity curve is. It is close to free for a steady, low-frequency configuration and severe for a variance-seeking one. This matters more than a pricing footnote, because the configurations that pass evaluations most reliably are precisely the lumpy ones — §5.3 shows that sizing up is how a zero-skill participant manufactures a pass rate. The consistency rule is thus positioned to penalise, at the payout gate, the same behaviour the evaluation gate rewards. That is the frequency asymmetry of §6 appearing in a second form, and it is the reason no single configuration is optimal across the account lifecycle.

### 3.5 Headline price is not cost

The quantity that enters the expected-value calculation is not the advertised fee but the total cost of reaching a funded account. Four components separate them.

1. Activation fees, charged separately on reaching funded status and frequently exceeding the headline fee by a large multiple.
2. Reset fees, charged to retry a failed evaluation, typically at or slightly below the original price.
3. Failed attempts. Since only a fraction of evaluations pass, the cost per funded account is the fee divided by the pass rate, not the fee.
4. Payout ceilings, which cap what a successful account can extract and therefore cap the return side of the same ratio.

The starkest instance is an account advertised at $39 — by a wide margin the cheapest entry point observed, and roughly a third the price of its nearest competitor. On reaching funded status it charges an activation fee of $189 at the 50K tier, giving a true cost to funded of $228. The account presented as the cheapest is in fact the most expensive of the three studied, by a factor of more than two. It also carries a 40% funded consistency rule, where the $98 account carries none.

| Account (50K tier) | Headline | Activation | Cost on funding | Funded consistency | Daily loss limit |
|---|---|---|---|---|---|
| Lucid Flex, evaluation | $98 | none | $98 | None | None |
| Tradeify, evaluation | $87 | none | $87 | 20% cap | $1,250 |
| TopOne Elite Access, evaluation | $39 | $189 | $228 | 40% cap | None |
| Lucid Direct, instant | $364 | none | $364 | 20% cap | $1,200 |
| TopOne InstantSim, instant | $407 | none | $407 | 20% cap | $1,250 |

*Table 4. Headline price against direct cost incurred on reaching funded status, 50K tier, before failed attempts. The cheapest headline is the most expensive account. Source: firm rules reference, 2026-07-10.*

A further constraint bounds the enterprise as a whole. Each firm caps the number of accounts a participant may hold, at five to ten depending on the account type. A participant cannot therefore scale a thin per-account expected value by repetition within one firm; the strategy must be replicated across firms, which multiplies exposure to unilateral rule changes and to discretionary non-payment. We return to this in §13.

### 3.6 The expected-value model

Bringing the components together, the quantity a participant should evaluate before purchase is:

$$EV_per_account = (pass_rate \times payout_rate \times payout_size) - cost_to_funded$$

Four terms, each of which must be measured rather than assumed, and each of which has a characteristic way of being got wrong.

1. Pass rate: the proportion of evaluation attempts reaching the target before the maximum-loss level, under the firm's real rules and using an evaluation-stage configuration. Not the funded-stage configuration; the two are different games.
2. Payout rate: the proportion of funded accounts that survive to and satisfy the payout window, including the minimum-days requirement and the consistency rule. The consistency-adjusted figure is the real one; the raw target-hit figure overstates it, in the 20% regime by roughly a factor of two.
3. Payout size: the dollars actually extracted per successful payout, capped at the firm's split and per-cycle ceiling. An uncapped equity figure carried through this term produced a false firm-level positive in early internal work, and the ceiling is not optional.
4. Cost to funded: total cost including resets, activation, and failed attempts, per §3.5.

**A property of this expression drives much of the paper: in the paper's primary comparisons the strategy enters principally through the first two terms, and weakly.** The account rules set the geometry, the price sets the hurdle, and the payout ceiling sets the prize. Table 12 makes the relative weights concrete: moving from the most to the least favourable account at a fixed win rate shifts expected value by more than the entire range of win rates plausibly available to a participant shifts it on a fixed account. The contract dominates the strategy, and that is the reason this paper analyses contracts rather than strategies, and it is also why a positive expected value here would be a statement about a business rather than evidence of a trading edge — a distinction §5.3 shows is not merely pedantic, because a zero-skill participant can manufacture a respectable pass rate by sizing alone.

### 3.7 How the surface shaped the research design

The rule surface was compiled before the substantive research began and determined the investigation in three ways, which is why the eventual null is a claim about the product rather than about our persistence. It fixed the account under study: the 50K tier with no consistency rule, no daily loss limit and end-of-day trailing is the most favourable structure among the three firms studied in detail, so a negative result there is conservative against every other row of Table 4. It fixed the two-stage structure of every simulation, since once the stages are understood as separate games with opposite optimal cadences, single-configuration modelling is indefensible. And it made the pass rate an object of measurement rather than an input.

**The third point is what changed the project.** Published evaluation pass rates — surveyed in §9.2 — run from 10.4% to 20.4% at futures firms on an account basis, while the practitioner calculator that popularised this business model runs on an assumed 33 percent. That gap changes the sign of the result and cannot be resolved by choosing one. At a \$98 fee and a 10 to 20 percent pass rate, cost per funded account is \$490 to \$980 against a modelled payout ceiling of \$1,000 — so on the published figures, a funded account that pays out with certainty barely clears the cost of reaching it. That observation turned the project from a search for a configuration into an investigation of whether the product admits a positive expectation at all.

## 4. Formal model

### 4.1 Setup

Let per-trade net profit and loss be independent and identically distributed as $N(\mu, \sigma^2)$, net of transaction costs. A participant executes $f$ trades per trading day, so daily profit and loss is $N(f\mu, f\sigma^2)$. Write $T$ for the profit target and $d$ for the maximum permitted drawdown. Equity is monitored at end of day. The evaluation terminates in success at the first session close with cumulative profit at or above $T$, and in failure at the first session close at or below the trailing loss level, which is (running end-of-day peak $-$ $d$).

Throughout, $T$ = \$3,000 and $d$ = \$2,000, with a ninety trading-day horizon.

### 4.2 The first-passage factor

The barrier components of the pass and payout probabilities of §7 are governed by the same first-passage law. We state it, and record a correction to an earlier internal draft.

**Proposition 3 (first-passage factor).** For an arithmetic Brownian motion with drift $\mu > 0$ and volatility $\sigma$, the probability that the process ever falls a distance d below its starting level is

$$P( \inf_t (\mu t + \sigma W_t) \leq -d ) = \exp(-2\mu d / \sigma^2), \quad d > 0$$

This is the standard one-sided first-passage result for drifted Brownian motion: the running minimum of such a process is exponentially distributed. Bailey and López de Prado (2013) apply the same machinery to drawdown-based stop-outs, which is the closest published treatment of the constraint studied here. The discrete random walk of §4.1 converges to it under diffusive scaling, and the exponential factor is the continuous-limit statement of the discrete gambler's-ruin ratio.

**Correction.** An earlier internal draft labelled $\exp(-2\mu d/\sigma^2)$ as the probability of a drawdown from the running peak. It is not; it is the probability of first passage to a fixed level d below the start. The two quantities differ, the peak-relative drawdown being the larger hazard, so the fixed-level factor understates the true risk faced by an account whose barrier ratchets. The no-go conclusion is invariant to the correction, since both bound the pass probability away from one for small $\mu$, and using the smaller factor is conservative for a negative result. §4.3 quantifies the gap.

### 4.3 The ratchet penalty

Proposition 3 records that the peak-relative hazard exceeds the fixed-level one but does not say by how much. The gap is worth measuring, because it is the quantitative content of the trailing rule — the specific parameter a firm can adjust to tighten the gate without changing the advertised target or drawdown. The comparison baseline is the two-barrier gambler's-ruin probability for the same process with fixed absorbing levels at +T and −d, P(hit +T before −d) = (e^{θd} − 1)/(e^{θd} − e^{−θT}) with θ = 2m/s², which is exact for a barrier that does not move. Comparing it against direct simulation of the ratcheting barrier at representative cells:

| Cell (μ, σ, f) | Fixed barrier (analytic) | Trailing DD (simulated) | Ratchet penalty |
|---|---|---|---|
| (5, 100, 10) | 0.871 | 0.731 | −14.0 pp |
| (10, 150, 5) | 0.841 | 0.691 | −15.0 pp |
| (2, 80, 20) | 0.746 | 0.580 | −16.6 pp |

*Table 5. The trailing-ratchet penalty: pass probability under a fixed barrier against the same account with a barrier that ratchets on the running peak. Source: reproduction run, 2026-08-31.*

Fourteen to seventeen percentage points of pass probability are consumed by the ratchet alone. This has a methodological consequence for the paper's framing: the closed form bounds, the simulation measures. The geometric scaffolding of the pincer is therefore an empirical regularity that the closed form constrains from above, rather than a derivation. That is deliberate and is not a weakening — a bound that is loose by fifteen points in the participant's favour and still yields a no-go is stronger than an exact calculation would be.

## 5. Simulation study: the (μ, σ, f) phase diagram under IID-normal profit and loss

### 5.1 Method

Four thousand Monte Carlo paths per grid cell, and two thousand five hundred in the sizing search. Daily loss limits, profit caps, minimum hold times, and consistency rules are excluded from the baseline. Each of those constraints only shrinks the feasible region, so the baseline is the generous case and the no-go statements derived from it in this section are conservative. Simulated trailing-drawdown pass rates are checked against the closed-form two-barrier bound of §4.3 and sit strictly below it, as they must.

## 5.2 Result 1: the required edge, by cadence

Taking the 0.50 pass-probability contour at moderate sizing (σ between roughly \$90 and \$130), the required net per-trade edge is:

| Trades per day | Net μ required for P(pass) ≈ 0.5 |
|---|---|
| 1 | \$31–34 |
| 5 | \$5.4–5.8 |
| 20 | \$2.0–3.4 |

*Table 6. Net per-trade edge required for a coin-flip pass probability. Source: phase-diagram simulation, this paper.*

> *These figures were re-derived for this draft and correct an error in the earlier tabulation, which reported \$25–30 at one trade per day. The original grid over μ was non-uniform (0, 1, 2, 3, 5, 8, 12, 20, 30, 50) but was rendered on a uniform index extent, so contour positions read off the figure were distorted. At σ = 90 the measured pass probability is 0.400 at μ = \$30, placing the 0.50 contour above rather than below that grid point. The corrected band is obtained by bisection on μ rather than by reading a contour, and Figure 1 is regenerated on a uniform grid. The error ran against the participant, so no downstream conclusion changes.*

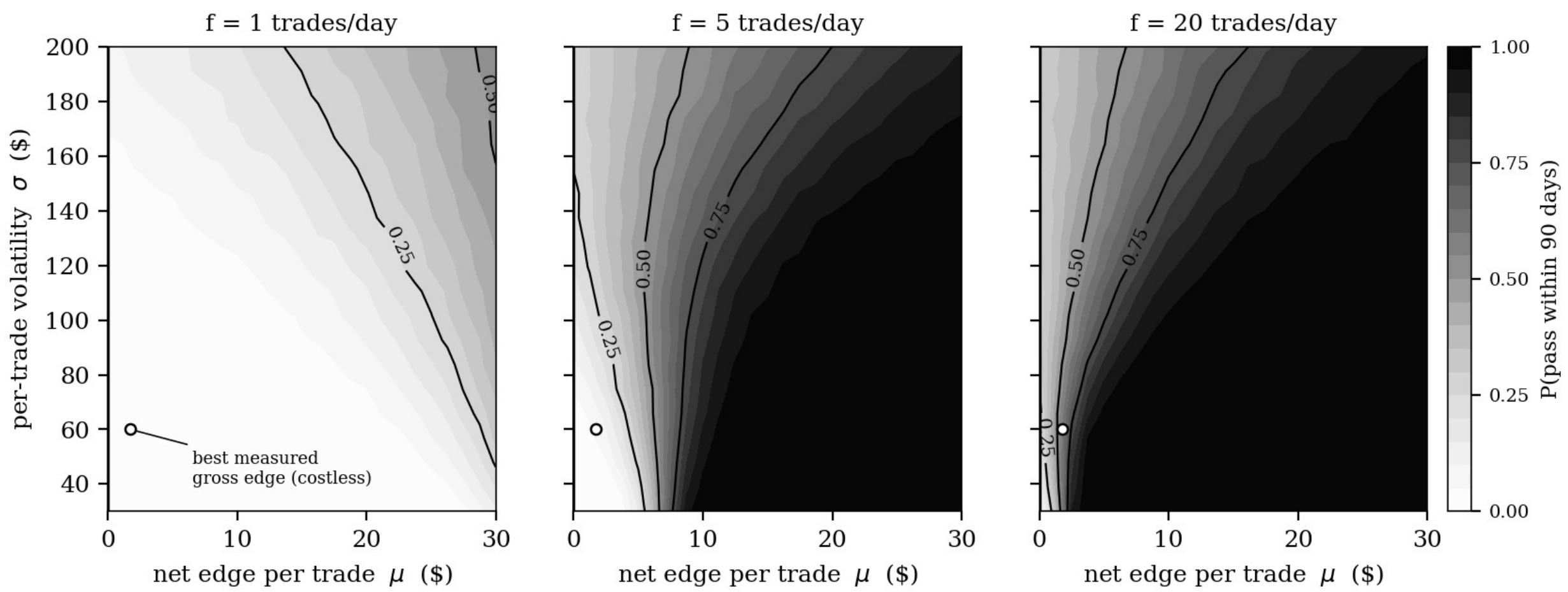

*Figure 1. Pass probability over the (μ, σ) surface at one, five, and twenty trades per day. Contours mark the 0.25, 0.50, and 0.75 levels. The vertical line at μ = 0 is the cost floor: once measured transaction costs are applied, every candidate in this research program sits on or to the left of it. The open marker is the best gross per-trade edge ever measured in the program (\$1.74 at σ = 60), plotted as if costs were zero.*

These are net figures. Adding the measured cost floor gives the required gross edge: on the full-size contract at five trades per day, gross μ of approximately \$20 to \$22 per trade, sustained. For calibration, the single best in-sample candidate measured across the whole programme — a market-on-close auction strategy — showed +\$16.91 per trade in sample and collapsed to −\$171.30 out of sample. The best bar-level gross edge measured across the whole programme was \$1.74 per trade; the Cycle II geometry sweep of Appendix C, a subset, tops out at +0.32 points. No measured candidate approaches the gross requirement at any frequency. Net of costs, every measured candidate sits at μ ≤ 0, which is off the left edge of the diagram entirely.

## 5.3 Result 2: pass rates are manufacturable by variance

Fix per-trade skill q = μ/σ and optimise over sizing σ. The critical feature of the resulting frontier is its left edge. At q = 0 — exactly zero skill — sizing-optimised pass probability is 0.36 to 0.47 on the regenerated grid of Figure 2, increasing in f.

The mechanism is direct. Under end-of-day monitoring with unconstrained sizing, the evaluation degenerates toward a single coin flip: as daily volatility grows, one day's close either exceeds +T or breaches −d, intraday paths are never observed, and by symmetry the pass probability approaches 0.5 independently of skill. The continuous-monitoring benchmark $d/(d + T) = 0.40$; discrete end-of-day observation with overshoot pushes above it. Adding a $1,000 daily loss limit, checked at end of day and terminal on breach, leaves the zero-skill optimum at 0.25 to 0.46.

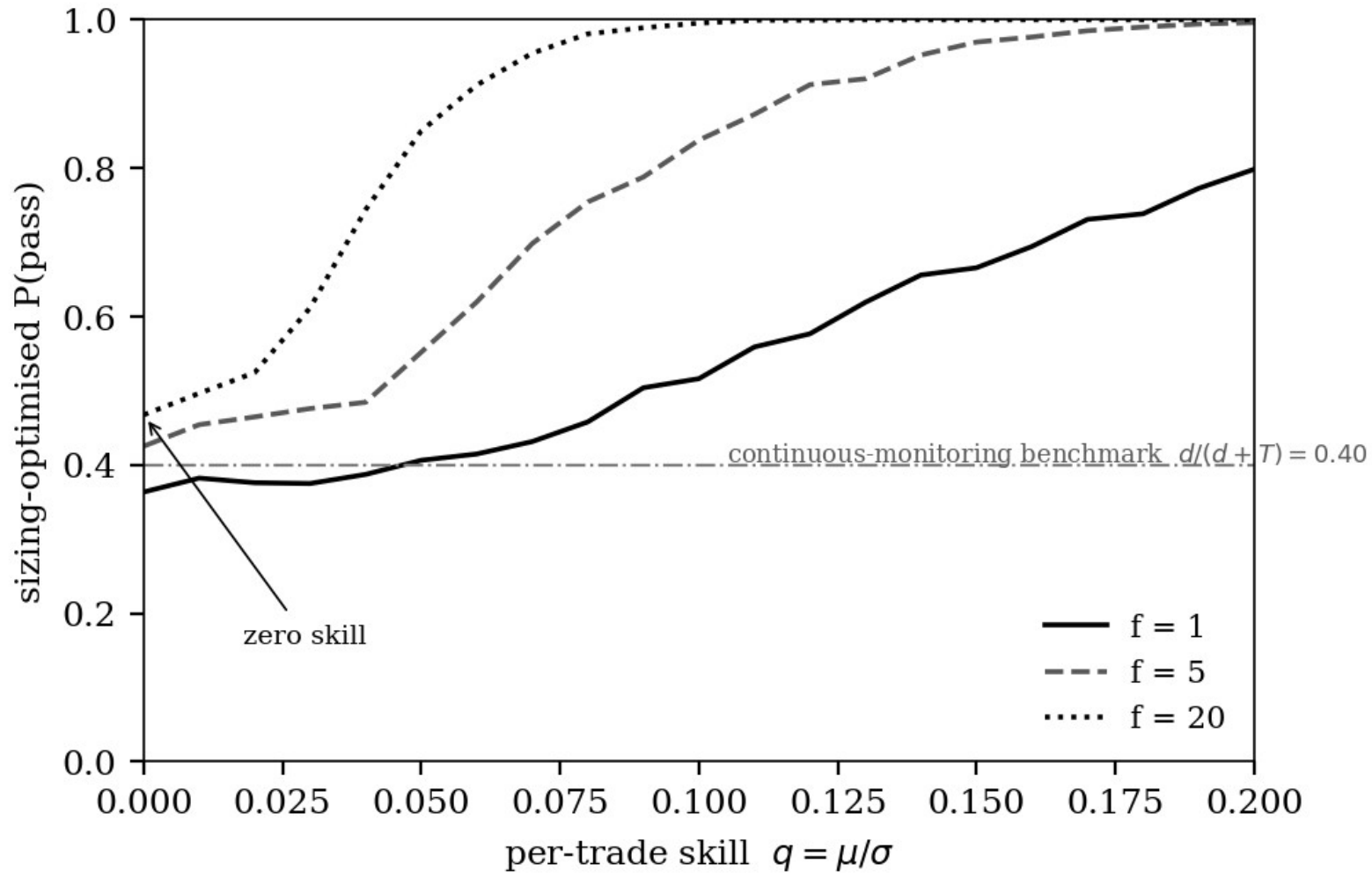

*Figure 2. Sizing-optimised pass probability against per-trade skill $q = \mu/\sigma$. The left edge is the result: at exactly zero skill, optimising over position size alone yields a pass probability of 0.36 to 0.47, rising in trade frequency. The dash-dotted line is the continuous-monitoring benchmark $d/(d+T) = 0.40$; discrete end-of-day observation with overshoot pushes above it.*

**A reconciliation is owed here, because §9.1 reports an observed evaluation pass rate of 16.8% and this section asserts 0.40 at zero skill.** Both are correct and they measure different objects. The figure above is the maximum over sizing: it describes a participant who chooses position size to exploit the barrier geometry and does nothing else. It is a statement about what the rules permit, not about what a population does. An observed cohort differs in two ways that both reduce the realised rate. Its members carry negative drift net of transaction costs rather than the zero drift assumed here, and §5.2 shows the required net edge is well above zero. And they do not size to the optimum — the sizing that maximises pass probability is aggressive, unintuitive, and directly contrary to the risk discipline the product's marketing encourages.

The gap between 0.40 and 0.168 is therefore the paper's claim rather than an embarrassment to it: the geometry permits a pass rate near four in ten to a participant with no skill whatsoever, and the observed population achieves fewer than two in ten. A skill-free construction outperforms the cohort, so the pass rate is not a clean measure of what the product claims it measures, and the evaluation is not merely easy either. What separates the two figures is position sizing, which is why §9.3 finds that roughly seventy percent of failures come from breaching a drawdown limit rather than from missing a profit target. The population is failing on the parameter that the zero-skill construction optimises.

**This is an instance of a known result. Goetzmann, Ingersoll, Spiegel and Welch (2007) show that when a principal judges an agent by a performance measure, the agent can game it by reshaping the return distribution, and that the gaming survives high transaction costs; an evaluation pass rate is such a measure and position size is the instrument. What is specific here is the magnitude, and it is the paper's pivotal structural claim.** If a participant with no edge whatsoever can manufacture a pass rate near 40% by sizing alone, then a pass rate cannot by itself be read as evidence of skill. This is weaker than saying the evaluation carries no information about skill — a noisy filter is still a filter, and §5.4 tests that question directly, finding that P(pass | skill) rises with skill. What it does establish is that any economically consequential selection not explained by the pass rate must occur downstream, at the payout gate. That is precisely what the measured joint hit rate shows: pass rate multiplied by payout rate lands in the range of one to three percent. The fee-insensitivity of negative expected value at measured drifts is consequently not an empirical curiosity but a forced consequence of the structure. This section derives the premise that §11 measures.

### 5.4 The skill gradient

Whether the evaluation carries any skill information is a separate question from whether a pass rate is informative, and it can be answered directly: simulate P(pass | skill) across a range of per-trade edges and see whether the curve is flat. It is not. At every cadence, pass probability rises monotonically in $q = \mu/\sigma$. The evaluation is a filter. The question is how strong a filter, and against what.

| $q = \mu/\sigma$ | f = 1, optimised | f = 5, optimised | f = 1, fixed σ = 90 | f = 5, fixed σ = 90 | f = 20, fixed σ = 90 |
|---|---|---|---|---|---|
| −0.04 | 0.322 | 0.378 | 0.000 | 0.021 | 0.035 |
| 0.00 | 0.370 | 0.434 | 0.000 | 0.101 | 0.263 |
| 0.02 | 0.373 | 0.449 | 0.001 | 0.190 | 0.488 |
| 0.05 | 0.409 | 0.554 | 0.002 | 0.394 | 0.774 |
| 0.08 | 0.449 | 0.749 | 0.004 | 0.630 | 0.928 |
| 0.12 | 0.578 | 0.914 | 0.011 | 0.876 | 0.987 |

*Table 7. Pass probability against per-trade skill, for a participant who optimises position size and one who holds σ fixed. Source: skill-gradient simulation, this paper.*

**The pass rate confounds skill with sizing at roughly equal weight.** At five trades a day and fixed sizing, moving from zero skill to a genuine edge of q = 0.05 raises the pass probability from 0.101 to 0.394 — 29 points. Moving from fixed to optimised sizing at zero skill raises it from 0.101 to 0.434 — 33 points. A participant with no edge who sizes aggressively is indistinguishable, by pass rate, from a participant with a real edge who sizes conservatively. This is the precise form of the claim: not that the evaluation filters nothing, but that what it filters cannot be separated from what it rewards. Two further features of Table 7 bear on the rest of the paper. At one trade a day with fixed sizing, pass probability is effectively zero at every skill level — the gate is impassable at low cadence without sizing up, which is §6's asymmetry from the participant's side. And the filter steepens sharply with cadence: at twenty trades a day skill separates participants far more than at five, which is the regime the funded stage then punishes.

### 5.5 Result 3: the cost–cadence pincer, generalised

Define expected calendar time per cleared account as expected days per attempt divided by pass probability, including failed attempts. The surface then reads as follows.

| μ (net $/trade) | σ | f | P(pass) | Days per clear | Years |
|---|---|---|---|---|---|
| $0.50 | 40 | 10 | 0.028 | 3,139 | ≈12.5 |
| $1.00 | 60 | 5 | 0.037 | 2,323 | ≈9.2 |
| $1.74 | 60 | 5 | 0.063 | 1,376 | ≈5.5 |
| $2.00 | 60 | 5 | 0.074 | 1,171 | ≈4.6 |
| $3.00 | 90 | 5 | 0.274 | 264 | ≈1.05 |
| $5.00 | 90 | 5 | 0.439 | 162 | ≈0.64 |
| $8.00 | 130 | 5 | 0.628 | 75 | ≈0.30 |

*Table 8. Expected calendar time per cleared $3,000 account. The $1.74 row grants the best gross bar-level edge ever measured in this program as if costs were zero. Source: phase-diagram simulation, this paper.*

The third row is the one that carries the argument. It grants the participant the best gross edge the program ever measured, and then grants it for free by setting costs to zero. Even under that double concession, clearing one $3,000 account takes approximately five and a half years at that sizing. With actual costs applied the row does not exist, because μ is negative. The earlier draft made this point with a single worked example at monthly cadence yielding roughly 11.4 years; the surface generalises it. The pincer is the region between a cost floor

pushing required gross μ upward and a cadence requirement pushing required pass probability and speed upward, and the measured strategy universe lies entirely outside it.

### 5.6 Caveats and robustness

1. Per-trade profit and loss is modelled as independent and normal in the baseline. Table 9 re-runs representative cells under three alternatives: Student-t with three degrees of freedom, a Bernoulli win/loss draw at 1:1.5 matched in mean and variance, and a persistent two-state volatility regime alternating between 0.6σ and 1.6σ. The normal, Student-t and Bernoulli results agree within about four points at every cell. Regime switching moves cells by −7 to +5 points with no consistent direction — clustered volatility helps at low sizing, where it makes the target reachable, and hurts at high sizing and cadence, where it clusters drawdown breaches. The cells nearest the required-edge contour of Table 6 — μ of \$3 and \$5 at σ = 90 — are stable under every model, and the sign of the no-go holds under each.

| (μ, σ, f) | Normal | Student-t(3) | Bernoulli 1:1.5 | Vol regime |
|---|---|---|---|---|
| (0, 90, 5) | 0.100 | 0.091 | 0.098 | 0.148 |
| (0, 300, 5) | 0.294 | 0.297 | 0.337 | 0.311 |
| (3, 90, 5) | 0.265 | 0.252 | 0.275 | 0.303 |
| (5, 90, 5) | 0.437 | 0.423 | 0.442 | 0.449 |
| (8, 130, 5) | 0.640 | 0.652 | 0.629 | 0.572 |
| (2, 60, 20) | 0.658 | 0.653 | 0.657 | 0.592 |

*Table 9. Pass probability under four per-trade distributions at representative cells, 6,000 paths each. Source: robustness simulation, this paper.*

2. The daily loss limit is modelled at end of day. Intraday marking strictly tightens it.
3. No consistency rules or profit caps appear in the baseline. Their inclusion only shrinks the feasible region, so the direction of the omission is known.
4. End-of-day trailing matches the account class studied. Intraday trailing on unrealised profit and loss is strictly harsher.
5. Monte Carlo standard error at four thousand paths is approximately ±0.8 percentage points at P = 0.5; contour irregularity in Figure 1 is sampling noise.

## 6. The frequency asymmetry

The two stages demand opposite trade cadences, and the size of the effect is large enough to dominate strategy selection. This section is the paper's cleanest mechanical finding, in the sense that it explains an observed industry practice rather than merely restating one.

### 6.1 In the evaluation, frequency is nearly free

| Trades per day | Pass rate at 40% win | Days to resolve |
|---|---|---|
| 1 | 23.4% | 11.4 |
| 5 | 24.9% | 3.6 |
| 20 | 25.4% | 1.6 |
| 30 | 25.0% | 1.4 |

*Table 10. Evaluation-stage pass rate and time to resolution by cadence. Source: two-stage simulation, this paper.*

Because the maximum-loss level is static within a session under end-of-day trailing, additional intraday trades do not compound drawdown risk within that session. Pass rate is essentially flat in frequency while time to resolution falls more than eightfold across the range shown. Within the modelled environment, greater cadence reduces expected time to resolution without materially reducing pass probability.

### 6.2 In the funded account, frequency is catastrophic

| Trades per day | Payout rate at 40% win |
|---|---|
| 1 | 34.8% |
| 3 | 29.9% |
| 5 | 23.4% |
| 10 | 11.7% |
| 20 | 3.9% |

*Table 11. Funded-stage payout rate by cadence. Source: two-stage simulation, this paper.*

The funded account must survive a five-day minimum. Twenty trades a day is a hundred opportunities to touch the floor. Within the modelled environment, the funded account is best traded as slowly as its minimum-day requirement allows.

### 6.3 The joint effect

Running twenty trades per day at both stages produces a joint gate of 1.0% at break-even. Splitting the cadence — fast in evaluation, slow when funded — produces **8.8%**. That is a nine-fold change in the joint gate induced by stage-specific cadence under otherwise identical strategy parameters — the interaction of cadence with the barrier and minimum-day rules, not frequency in isolation.

This is exactly the split that practitioners describe and never explain. The mechanism is the interaction of the minimum-days survival requirement with end-of-day trailing. Note the implication for the paper's framing of the pincer: it is not eval-stage frequency against friction. It is that the two stages demand opposite cadences, so a participant running one strategy at one frequency is necessarily suboptimal at one of the two gates.

## 7. The break-even threshold

Expected value per account is the product of pass rate, payout rate, and payout size, less the total cost to funded. Costs are inclusive of evaluation fee, expected reset costs, and activation fees; payout size is capped at the firm's real per-cycle ceiling and split, modelled here at $1,000, being fifty percent of a $2,000 drawdown.

$$EV_per_account = (pass_rate \times payout_rate \times payout_size) - cost_per_account$$

### 7.1 Expected value by account

| Win rate | Lucid Flex 50K | Tradeify 50K | TopOne 50K |
|---|---|---|---|
| 40.0% | −$10 | −$43 | −$140 |
| 42.0% | +$42 | −$10 | −$88 |
| 44.0% | +$106 | +$38 | −$24 |
| 46.0% | +$181 | +$98 | +$51 |
| 48.0% | +$265 | +$170 | +$135 |

*Table 12. Expected value per evaluation by account structure and win rate at 1:1.5 reward-to-risk. Source: two-stage model with sourced pricing.*

On Lucid Flex — the most permissive rule geometry of the three, with no consistency rule and no daily loss limit — break-even lies between 40.9% and 41.5% across the two implementations reported below. The lowest threshold across all three accounts is Tradeify's 40.5% on the independent solve, the consistency rule barely binding at steady cadence as §3.4 explains. The driftless baseline at that reward-to-risk ratio is 40.0%, so the requirement is roughly 1.5 percentage points of genuine edge on the most permissive rule geometry, and half a point on the cheapest. The threshold rises to about 44% on 20%-consistency accounts at lumpy sizing and above 46% on high-fee accounts.

The threshold was re-derived on an independent implementation for this draft: a two-stage simulation with end-of-day trailing, costs applied per trade, a fast evaluation configuration and a slow funded configuration, solving directly for the win rate at which expected value crosses zero.

| Account | Break-even win rate | Over driftless 40.0% |
|---|---|---|
| Lucid Flex (no consistency rule) | 40.9% | +0.9 pp |
| Tradeify (20% cap) | 40.5% | +0.5 pp |
| TopOne Elite Access (40% cap) | 43.6% | +3.6 pp |

*Table 13. Break-even win rate solved independently. Source: independent two-stage solver, this paper.*

**The independent solve returns 40.9% on Lucid Flex against the 41.5% reported above — agreement to within modelling choices, and confirmation that the threshold sits within about a point of the driftless baseline.** We record this because two larger figures have circulated in the underlying research notes, expressed as edges of roughly four and six percentage points over a measured strategy baseline. Neither is a break-even. The first is a working bar and the second adds a margin buffer for variance and rule changes; both are business thresholds for deciding whether to commit capital, not statistical statements about where expectation crosses zero. Quoting them as break-even overstates the requirement by three to five points of win rate. This paper reports break-even only, always as an absolute win rate at 1:1.5 with the driftless 40.0% given as reference, and never as an increment over a strategy-specific baseline.

### 7.2 Restating the no-go

Earlier drafts stated the gate compression as a single structural constant of approximately 2.5%. That overstates what the data supports. The figure decomposes into three separate contributions: a measured win rate slightly below break-even; the minimum-days gate, which is independent of drawdown geometry and accounts for roughly half the compression; and modelling choices that were pessimistic relative to real firm rules, specifically intraday rather than end-of-day trailing and single rather than split cadence.

The defensible form of the claim is therefore conditional and account-specific, stated at the top of the implementation range: *the evaluation product is negative in expectation for any participant whose win rate at 1:1.5 falls below approximately 41.5% net of costs on the most permissive rule geometry available, and below 40.5% on the cheapest; the threshold rises to approximately 44% on 20%-consistency accounts at lumpy sizing and above 46% on high-fee accounts.*

This is a stronger paper than an unconditional claim would be. It is falsifiable, it is account-specific, and it survives the obvious referee objection that the author selected pessimistic parameters — because the stated threshold is derived on the account structure most favourable to the participant, and the source record documents each modelling correction and its direction.

## 8. The kill record

The preceding sections establish a threshold; this section reports what was measured against it. Three research cycles over roughly two dozen mechanism-first candidates, followed by two account-level measurements. Every candidate is reported, including those that passed certification and were later withdrawn, because a kill record is only evidence if it includes the near-misses.

| Cycle | Question | Candidates | Survivors |
|---|---|---|---|
| I | Is there a bar-level intraday edge tied to forced flow? | ~22 | None advanced |

| Cycle | Question | Candidates | Survivors |
|---|---|---|---|
| II | Can the evaluation be beaten as a churn business? | 1 strategy + 10 geometries | None |
| III | Does forced/informed separation appear at tape resolution? | 2 funded, 2 deferred | None |

*Table 14. The three research cycles. Source: research ledger, this programme.*

### 8.1 Cycle I — the bar-level edge search

The first cycle sought an intraday reversal or continuation tied to a named forced counterparty, on one-minute bars. Index reconstitution and quarterly-roll settlement flow, and options-dealer gamma hedging around large open-interest strikes, were both killed with no net-positive out-of-sample result at any tested geometry. Liquidation-cascade reversion was deferred for finer data and became the Cycle III programme. Across all bar-level strategies, zero net-positive out-of-sample results over 2021 to 2026.

A second batch of eight mechanisms was specified and swept. Six were killed outright, one of them explicitly for inability to clear the round-turn friction that had killed its coarser predecessor. One was killed on both legs after an authorised per-expiry data pull. The eighth is the cycle's most instructive result and is reported in full.

**The withdrawn certification.** A gamma-pin reversion candidate passed every gate in sequence: positive net-of-cost edge, clearing the Sharpe floor, formally certified by the risk function and escalated for a paper-trading decision as the cycle's first full-gate pass. Within the same cycle it was retracted. A closer out-of-sample consistency check showed the certified edge did not hold on the held-out sample, and the escalation was withdrawn before any capital decision was taken. The cycle produced zero eligible candidates. That a candidate could pass certification and still be killed on a second, stricter look — rather than advanced because it had already cleared the bar — is the behaviour that makes the eventual null credible. The programme's failures were not for want of candidates that looked ready.

A third cycle ran four waves of ten mechanism-first candidates. Two cleared the risk ladder and are the closest the programme came to a survivor.

| Candidate | Mechanism | IS → OOS Sharpe | OOS n | Disposition |
|---|---|---|---|---|
| W1-01 | LETF end-of-day rebalance, long | 0.70 | 74 | Killed, Sharpe below floor |
| W1-06 | LETF EOD convexity, short side | 1.91 → 3.07 | 60 | Certified, escalated, not advanced |
| W1-02 | Clears ladder mechanically | 0.14 IS | — | Rejected: OOS is 2024 only; 2025 and 2026 negative; 44% overlap with W1-06 |
| C10 | Holiday-eve forced de-risk | → 3.15 | 53 | Near-survivor; matched control pending |

*Table 15. The two near-survivors and the two candidates adjacent to them. Source: research ledger, Cycle III waves.*

Neither advanced. W1-06 rested on a single mechanism over roughly a dozen large-down days per year. C10's surviving signal was momentum rather than the pre-specified forced-sell, so its decisive test was a matched control — generic close momentum on all sessions against the pre-holiday subset — with a standing instruction to kill if the edge was not specific to the forced-flow condition. The lesson carries into §10: an out-of-sample Sharpe above the floor on a narrow event set is not a durable edge, and it is not an edge an evaluation account can use.

### 8.2 Cycle II — the churn thesis

If no single edge survives, perhaps the evaluation can be beaten as a game: buy cheap accounts, pass some by variance, extract faster than accounts cost. This is the intuition §12 formalises as the hedged pair. Two tracks were tested.

The first track tested a mechanical fair-price reversion around the New York open. Its first pass looked marginally positive: a twenty-contract morning-only configuration reported a 28.5% out-of-sample pass rate and an extraction expected value of +$362 per account. The per-firm split shows that positive appearance was entirely an artefact of the firms whose constraints had not yet been modelled.

| Account | OOS EV per account | Note |
|---|---|---|
| Apex | +$469 | Intraday trailing not yet modelled |
| Fifth firm | +$469 | Intraday trailing not yet modelled |
| Generic instant 50K | +$362 | Headline figure; 72.3% account-blow rate |
| Lucid Flex | +$226 | Payout eligibility collapses to 14.5% under a 50% consistency rule |
| Sixth firm | −$77 | Five winning days plus 50% consistency kills a fast, lumpy configuration |

*Table 16. First-pass per-firm expected value, before correction. Every positive cell belonged to a firm whose constraints had not yet been modelled. Source: research ledger, Cycle II Track A.*

The re-run under corrected firm constraints killed it. At the measured $1,200 daily loss limit the per-trade gross edge was between −0.08 and +0.22 points and sign-unstable. The break-even payout required on the instant account was $27,921 — cost divided by a 0.5% eligible pass rate — against a measured profit-at-pass of roughly $3,700. The blow rate was 78.9% in sample and 71.5% out of sample at the sizing the winners required. The decisive constraint was the $2,000 trailing drawdown rather than the daily loss limit: moving the personal risk cap shifted the blow point by fewer than five index points.

**The reinstatement challenge.** The kill was not accepted on first statement. Because a strategy killed under the wrong drawdown model would be a false kill, an explicit reinstatement test was issued: state exactly what the strategy was killed on — the gate, the measured value, the threshold — and confirm the kill holds under real firm rules; if the kill assumed a drawdown model no real firm uses, the kill is void and the strategy is reinstated. The challenge was material, because the strategy had measured a 28.5% out-of-sample pass rate with no in-sample to out-of-sample decay, which is not the profile of a spurious result. The resolution held the kill: under end-of-day trailing modelled correctly, the payout-rate verdict of §8.4 still placed the best committed rate below its break-even line, and the negative net-of-cost per-trade edge is firm-independent and evaluated before any payout term enters. We report the episode because a null that survived a good-faith attempt at revival is stronger evidence than one that was never challenged.

The second track swept ten mechanical geometries. Best gross edge +0.32 points per trade; eight of ten gross-negative; net of costs all ten have negative expected value per trade (Appendix C). The initial run reported survivors, all artefacts of a missing account-survival term: auditing its thirty-five positive-EV cells under a drawdown-first gate found zero survivors, blow rates of 62.8% to 79.1%, and thirty-four of thirty-five at the maximum twenty contracts. An independent re-run cleared none of sixty; a later gate none of seventy-two.

**Two ranking defects were found and corrected mid-programme, and both are reported because they are instructive about the direction of error.** A geometry-search script imported the attempt simulator but never the account-survival term, manufacturing apparent survivors that vanished under a proper ruin gate. Separately, an extraction-value ranking key proved ruin-seeking: it was maximised by maximum leverage, and a zero-edge coin flip scored +$721 per attempt under it. A third defect was found in the simulator itself, which checked the drawdown floor only at trade close and so never saw intrabar adverse excursion; a maximum-adverse-excursion accounting blew 74.80% of attempts against the engine's 71.52% on the same cell. Every one of these corrections moved results toward the null and never away from it. The engine was the optimistic instrument throughout, which is why the measured result should be read as an upper bound on participant expectation rather than a point estimate.

### 8.3 Cycle III — order flow at tape resolution

The final cycle asked whether the separation between forced flow, which mean-reverts, and informed flow, which continues, becomes tradable at tick resolution. Two candidates were funded from a fixed $250 data budget. Both are dead.

| Candidate | In-sample | Out-of-sample | Verdict |
|---|---|---|---|
| Sweep / liquidation cascade fade | −$9.49/RT, Sharpe −0.222, 43.6% win, n=1,786 | −$27.61/RT, Sharpe −0.646, 42.3% win, n=6,236 | Kill: cascades continue 57% of the time |
| Closing-auction pressure | +$16.91/RT, Sharpe +1.011, 57.1% win | −$171.30/RT, Sharpe −12.097, 16.7% win, n=6 | Kill: in-sample edge fully reverses |

*Table 17. Cycle III, tick-resolution candidates. Source: research ledger, Cycle III.*

**The second row is the cautionary case of the entire programme.** A clean, positive in-sample result with a Sharpe just above one, which inverts completely out of sample on a sample far too small to gate — the signal fires in only seven to twelve percent of sessions, giving six out-of-sample trades against a thirty-trade minimum. It is the single best in-sample candidate the programme ever produced, and it is worse than useless. Any account of these results that reports in-sample figures is reporting this.

Two further candidates were deferred as unfundable within the data budget. Full-size contract economics of $13.80 per round turn dominate the sub-tick separation any of these could detect.

### 8.4 The payout-rate verdict

No candidate produces a positive net-of-cost edge. The remaining question is whether the account structure can be made to pay independent of edge. This measurement argmaxed the consistency-adjusted payout rate over reward-to-risk, position size and cap, in sample, with one out-of-sample validation.

| Firm | Committed (IS) rate | OOS rate | Coin flip | Blow rate | Break-even |
|---|---|---|---|---|---|
| Lucid Flex | 4.6% | 26.2% | 2.6% | 98% | 9.5% |
| TopOne Elite Access 50K | 2.1% | 10.9% | 2.6% | 98% | 10.2% |
| Tradeify | 1.3% | 11.5% | 2.6% | 98% | 9.5% |
| Lucid Flex (instant) | 1.3% | 11.5% | 2.6% | 98% | 12.5% |
| TopOne Elite Access (instant) | 1.3% | 11.5% | 2.6% | 98% | 14.0% |
| TopOne Elite Access (daily) | 2.1% | 5.1% | 2.6% | 100% | 15.9% |

*Table 18. Payout-rate optimisation by firm. Every row fails the durable-survival criterion. Source: payout-rate optimisation, this paper.*

Three findings settle the test, and they would have settled it in either direction.

1. The committed number fails. The best achievable in-sample payout rate is 4.6% at Lucid Flex, against a 9.5% break-even line. The number a participant can commit to in advance does not clear.
2. The high out-of-sample rates are a maximum-variance lottery rather than a business. Every winning cell is the most aggressive configuration available and breaches the floor approximately 98% of the time. The apparent payout is an account spiking to +$3,000 on variance and extracting once immediately before its near-certain failure. In-sample to out-of-sample the rate swings by a factor of three to six, which is a regime coin toss, not a stable input.
3. A zero-edge coin flip matches the committed rate. The control posts 2.6% at the winning configuration, inside the band of the strategy's committed rates of 1.3% to 4.6%. The measured payout rate is capturing the firm's option value and the position's variance, not any property of the strategy.

**The survival test is the decisive one.** The only configurations that survive — meaning a blow rate at or below 50% — pay out 0.0%. A durable account never reaches the withdrawal target. Every configuration that reaches the target blows 80% to 100% of the time. No corner of the parameter space both survives and extracts.

### 8.5 The adversarial spine

The obvious objection to test one is strategy selection: perhaps the strategies tested were simply poor. The second test is designed to close that objection by removing the author's discretion from strategy choice entirely.

We take the only fully-specified, zero-discretion rule set published by the most prominent practitioner in this space — a displacement-candle entry with stated continuation, stop, target, body-size and distance parameters — and implement it without interpretation. A 324-configuration in-sample sweep, selection in sample only, one out-of-sample validation, no re-selection, full costs, zero-edge control attached. Continuous front-month index futures at one-minute resolution: 2,832 in-sample sessions to September 2021, 1,217 out-of-sample to June 2026.

**Out-of-sample validation of the in-sample-selected configuration: n = 1,217, win rate 38.54%, mean −$32.14 per trade.** That is below the 40.0% driftless baseline at the selected reward-to-risk ratio. The published rule set has negative out-of-sample edge.

| Firm | | Pass | Payout | Break-even | Margin | EV/ account |
|---|---|---|---|---|---|---|
| Lucid Flex | Rule set | 15.25% | 16.16% | 21.42% | −5.26 pp | −$24.06 |
| | Coin | 20.50% | 12.50% | 15.93% | −3.43 pp | −$21.12 |
| Tradeify | Rule set | 15.25% | 16.16% | 19.56% | −3.39 pp | −$15.10 |
| | Coin | 20.50% | 12.50% | 14.55% | −2.05 pp | −$12.25 |
| TopOne Elite Access | Rule set | 15.25% | 16.16% | 48.89% | −32.73 pp | −$152.63 |
| | Coin | 20.50% | 12.50% | 36.37% | −23.87 pp | −$149.64 |

*Table 19. The published rule set against the zero-edge control, by account. Source: adversarial test, this paper.*

Three independent readings of the failure resolve to one cause.

1. Negative out-of-sample drift. A 38.5% win rate at 1:1.5 where 40% is a coin. Every account-level number downstream is the consequence of dragging negative drift through a trailing barrier.
2. The control out-passes the strategy. Coin pass rate 20.50% against the rule set's 15.25%. A driftless walk is near-optimal against a roughly symmetric barrier; a negative-drift walk is worse than random at the single thing the evaluation measures.
3. The joint gate equals the control's joint gate, to within 0.10 percentage points on all three firms. Whatever passes through these accounts is the account's option value. The strategy contributes nothing.

### 8.6 Convergence

The two tests used different entry mechanisms, different simulator builds, and were run several weeks apart by different means. They return the same verdict by the same mechanism. An independent adversarial implementation reproducing the earlier verdict is the strongest form this result can take. The in-sample to out-of-sample decay — 47.3% to 38.5% — also reproduces the pattern observed across the full research ledger, in which every in-sample-positive candidate in the program's history decayed out of sample.

The practitioner claims a 57.5% win rate on this rule set; the mechanical skeleton measures 38.5%. The nineteen-point discrepancy cannot be attributed from our data to the published rule set. It could reflect the discretionary overlay he states on camera that he withholds, or other implementation differences we cannot observe. He has himself said the skeleton alone would break even or lose; measured, it loses. What is established is narrow: the transferable, specifiable component of the published material has no edge.

# 9. External evidence

Section 8 reports what this programme measured. This section reports what the sector publishes, and compares the two. The engine behind §8 was calibrated against an independent commercial backtesting platform, agreeing to a profit factor of 0.970 against 0.957 over approximately 3,800 trades with identical trade endpoints — a check on execution mechanics rather than on outcomes. What follows is the check on outcomes.

## 9.1 A futures firm publishing cohort statistics

**The closest available comparison is a futures firm publishing full-year cohort statistics on its own site.** Its account is the archetype modelled throughout this paper: a 50K account with a $3,000 profit target, a $2,000 maximum loss limit on end-of-day trailing that locks once it reaches the starting balance, five qualifying days at a minimum daily profit, and a consistency requirement on the alternative payout path. The rule geometry is not merely comparable to §3; it is the same geometry.

| Published quantity, calendar year 2025 | Value | Denominator |
|---|---|---|
| Evaluations initiated that were completed | 16.8% | Accounts |
| Participants advancing to funded at least once | 51.8% | Individuals |
| Funded participants receiving a payout | 33.3% | Individuals |
| Funded participants reaching a live capital account | 0.71% | Individuals |

*Table 20. Futures firm cohort disclosure, January–December 2025. Published by the firm; denominators are as stated by the firm and are not uniform. Source: firm risk disclosure, retrieved 2026-08-31.*

**The payout rate is the clean comparison, and it lands almost exactly on the model.** This paper's two-stage simulation puts β at 34.8% for an end-of-day-trailing account with a five-day requirement and no binding consistency rule (Table 3, 40% win-rate row); the independent solver of §7 returns 34.6% at its break-even. Topstep reports 33.3% for calendar year 2025; a second futures firm, MyFundedFutures, publishes 28.6%. Our figures sit between the two and within six points of both, on comparable rule geometry, from populations the authors did not collect and could not influence. Two firm-published values rather than one is what the payout-rate claim now rests on.

The evaluation pass rate is not a clean comparison and we do not treat it as one. The firm reports 16.8% against our simulated 28–41%, but our figure is generated by a zero-drift participant optimising position size, which is a deliberately favourable construction and not a description of the population. A real cohort carries negative drift net of costs, and §5.3 predicts that the sizing-optimised figure is an upper bound on what an actual participant achieves. The direction of the discrepancy is the direction the model requires.

**The fourth row is the one that bears on §12.** Fewer than one funded participant in a hundred and forty is advanced to an account trading real capital. Whatever the evaluation is selecting for, it is not selecting traders to back: the pipeline to live capital is, by the firm's own published figure, essentially closed. This is the strongest available evidence against the talent-identification framing dismissed in §13.1, and it is the firm's number rather than ours.

The structural pattern of §12.7 also recurs. A fourth futures firm imposes a minimum reserve balance that cannot be withdrawn until the account terminates, which is the safety net of §12.7 under another name, together with a 40% consistency test applied at payout and an activation fee on passing. Three of the four futures firms examined here place their binding constraints in the funded account rather than the evaluation, and the fourth is the one whose evaluation is hardest.

Two structural details of the same disclosure confirm points made earlier from other firms. The evaluation is sold as a monthly subscription that rebills until the participant passes or cancels, which is the recurring-fee structure flagged in §12.6 and means the cost-to-funded of §3.5 accrues per month rather than per attempt. And passing incurs a separate activation fee of $149 on the standard path, which is the headline-versus-true-cost divergence of §3.5 appearing at a second firm.

### 9.2 What the sector publishes, and what it does not

We surveyed the sector for any published funnel statistic. The result is reported in full below, including the firms that publish nothing, because the pattern of disclosure is evidence in its own right.

| Source | Asset class | Pass rate π | Payout rate β | Basis |
|---|---|---|---|---|
| Topstep | Futures | 16.8% (accounts) 51.8% (persons) | 33.3% | Firm disclosure, full-year 2025 |
| Take Profit Trader | Futures | 20.4% | — | Firm reporting, Jan–Aug 2023 |
| MyFundedFutures | Futures | — | 28.6% | Firm-published, checked Aug 2026 |
| Earn2Trade | Futures | 10.4% | — | Verified figure, 2024 |
| Apex Trader Funding | Futures | 15–20% first attempt ~40% with resets | — | Firm claim, unaudited |
| FPFX Technology | Mostly FX/CFD | 14% (persons) | 45% (persons) | 300,000+ accounts, 100,000 traders, 10 firms; Finance Magnates (2024) |
| FTMO | FX | 9–10% | — | Firm citation, two-step |
| The Funded Trader | FX | 5–10% | ~20% | Firm disclosure, March 2025 |
| ATFunded | FX | ~6% overall | — | Firm social disclosure, June 2025 |
| Velotrade report | Mixed, 6 firms | — | 7% of buyers paid | Rulebook review, 2026 |

*Table 21. Every published funnel statistic we could locate; each source is listed in the References under the firm or organisation named. Dashes indicate the quantity is not published.*

**The column of dashes is the finding.** We identified more than twenty firms operating in this sector. Eight publish a pass rate in any form. Three firms publish a conditional payout rate themselves — two of them futures firms — and two cross-firm sources report buyer-level payout outcomes. Several publish cumulative payout totals — figures in the hundreds of millions — which are numerators without denominators and cannot be converted into a rate. The asymmetry is systematic: π is disclosed and marketed, β is disclosed rarely and marketed never.

That asymmetry is precisely what §13.3 predicts. The argument there is that π is cheaply observable and therefore useless as a source of profit, while β is realised only after the fee is sunk and is where the belief gap must live. A seller reasoning that way has every incentive to publish the first and none to publish the second. We did not construct this test either; the disclosure pattern was already in the world, and it matches.

### 9.3 Patterns across the survey

Three patterns survive the small sample.

**Futures pass rates exceed foreign-exchange pass rates, consistently.** The four futures figures span 10.4% to 20.4%; the foreign-exchange figures span roughly 6% to 10%. Futures firms run one-step evaluations with no time limit and, at the extreme documented in §12.6, no consistency rule and no minimum days; the foreign-exchange standard is two-step with a ten percent target. That comparison is confounded by asset class, firm and period, so we prefer a cleaner test one of the sources supplies internally. The Swiset study covers approximately ten thousand traders across eleven regions in a single window and reports its results split by evaluation architecture rather than by firm: one-phase evaluations pass at 20.3%, two-phase at 11.8%. Same population, same period, same methodology; the study does not control for firm composition, account size or pricing within each architecture, so this is a within-study association rather than an identified causal effect, but it moves the pass rate by a factor of roughly 1.7. The softer architecture is associated with the higher rate.

**The joint rate cannot be compared across sources without fixing the denominator, and doing so dissolves an apparent agreement.** FPFX Technology's dataset, reported by Finance Magnates (2024), gives 14% of traders passing and 45% of those receiving a payout, hence 7% of buyers paid — all per person. An independent rulebook review of six firms reports the same 7% of buyers, at the same denominator, which is a genuine corroboration. Topstep is not comparable to either without care: its 16.8% is per account, its 51.8% is per person, and its 33.3% payout rate is per person. The person-level joint implied by its own figures is 51.8% × 33.3% ≈ 17%, not the 7% of the cross-firm datasets. Multiplying its account-level pass rate by its person-level payout rate, as a careless reading would, yields 5.6% and a spurious agreement. We report the corrected figures and draw no conservation claim from them.

Three readings of the residual gap are available and we cannot distinguish them. Futures accounts may genuinely deliver a higher person-level joint rate than foreign-exchange accounts, which would be consistent with their softer evaluations. Topstep may be a favourable outlier, which its unusual willingness to publish would predict. Or the cross-firm datasets may count buyers in a way that includes abandoned or never-traded accounts, depressing their denominator. What survives is narrower than a conserved constant: at every denominator and every source, the share of participants who ever receive a payout runs from single digits to the high teens, and in no dataset does it approach the rate a participant would need for the product to pay.

Failure attribution is consistent and confirms the mechanical claim. One source attributes roughly seventy percent of failures to loss limits, decomposing as fifty percent maximum drawdown and twenty percent daily loss caps. A second, independently compiled, gives forty-five to fifty-five percent daily loss limit, twenty to thirty percent trailing drawdown, ten to fifteen percent consistency violation, and five to ten percent time expiry. The two disagree on the split between the drawdown types and agree that drawdown mechanics account for the large majority. Missing the profit target is not a leading cause of failure anywhere in the published data.

### 9.4 The seller's revenue base

A structural fact about the sector's economics bears directly on §13.1, and it comes from industry analysis rather than from us. Firms in this sector are reported to derive eighty to ninety-five percent of revenue from evaluation fees. If that is even approximately right, then the arrangement is not a trading business that funds itself from trader performance; it is a fee business whose product is the evaluation. The zero-sum identity $\Pi = -E$ of §13.1 is a statement about the contract, not an accounting of any firm; but if fees are eighty to ninety-five percent of revenue, the contract is most of the business, and the identity describes where the money comes from more closely than an abstraction usually does.

Two corollaries follow that are visible in the sector's history. A fee-funded business is vulnerable to any interruption in new purchases rather than to trading losses, and the sector experienced a contraction in 2024 in which an estimated eighty to one hundred firms ceased operating, roughly thirteen to fourteen percent of global operators, on a platform and regulatory shock rather than a market one. And the largest documented failure had collected approximately $310 million in fees before closing. Neither fact is an allegation about any surviving firm. Both are consistent with a business whose income is the fee stream and whose liabilities are the payouts, which is the structure §13 models.

**This is behaviour consistent with the mechanism of Proposition 5, in data we did not construct.** §13.2 argues that the seller optimises the product of the two factors rather than either one, because it is the product that determines $\Pi$ and the placement of friction between stages is otherwise free. If that is right, where friction is placed should matter more than the pass rate alone, and a harder evaluation need not imply a better-selected funded population. The available datasets are consistent with that weaker prediction: architecture materially affects pass rates, while separate payout datasets show that downstream payout incidence remains low. A model in which the evaluation is a genuine skill filter predicts something different — a harder filter should admit better traders and raise the downstream payout rate more than proportionally — and the cross-asset data does not show that. What appears is consistent with substitution, friction moved from one gate to the other. It does not establish a conserved joint rate across firms or denominators; §9.3 explains why that stronger claim, made in an earlier draft, does not survive.

### 9.5 Limits of the external evidence

**The payout rate is the paper's thinnest empirical footing and we do not disguise it.** Four values exist — 33.3% and 28.6% at futures firms, 45% and approximately 20% elsewhere — a range of more than two to one across asset classes, methods and years. The two futures figures are close to each other and to our simulated 34.6%, which is why the claim rests on them; the other two are context. Both publishing firms do so because they are unusually transparent, and if transparency correlates with favourable numbers their figures are upper bounds on the sector. That bias runs toward our conclusion, which is the safe direction but does not enlarge the sample.

Three further limits should be stated plainly, because the agreement above is easy to over-read.

1. The external sample is largely not futures. The instrument-specific cost law of §3.6 — fixed commission being an order of magnitude heavier in proportional terms on the micro contract — has no analogue in a spread-based foreign-exchange product. The joint-rate agreement is therefore evidence about gate design, which is common to both, and not about transaction costs, which are not.
2. Denominators are not uniform, and the firm disclosure of Table 20 states four figures on two different denominators. Its pass rate is per account; its payout rate is per individual. Multiplying them, as an earlier draft of this paper did, is therefore an approximation rather than a measured joint rate, and it understates the per-individual figure: the same disclosure reports that 51.8% of individuals eventually pass at least once, against 16.8% of accounts. A participant who buys ten evaluations is ten accounts and one person. We report both denominators rather than choosing the flattering one, and no argument in this paper rests on the multiplication.
3. Many firms also publish cumulative payout totals rather than rates. A firm reporting hundreds of millions in cumulative payouts is reporting a numerator without a denominator, and the quantity that matters here is the ratio. These figures are reproduced above only as disclosures, and no rate is inferred from them.

### 9.6 The regulatory position

The paper is careful throughout to claim negative expectation under published rules and not to allege fraud, and the regulatory record shows why that distinction is worth the discipline it costs.

The most substantial enforcement action brought against a firm in this sector, CFTC v. Traders Global Group Inc., alleged fraud exceeding three hundred million dollars arising from simulated trading. It was dismissed with prejudice on 13 May 2025 and the agency sanctioned, following a Special Master's report finding that the regulator had made false representations to the court and acted willfully and in bad faith; the court awarded the defendants over three million dollars in fees. Whatever that outcome establishes about the particular defendant, it leaves the sector without the precedent the action sought, and it is a demonstration that allegations of this kind are difficult to sustain. A structural argument from published rule geometry and measured costs does not depend on any such allegation and is not weakened by that dismissal.

The sector remains outside the regulatory perimeter in the jurisdiction that matters most for the firms studied here, and the position is moving in different directions on either side of the Atlantic. European authorities have converged on treating evaluation products as potential investment services: Belgium's FSMA issued a consumer warning on 7 March 2024, Italy's CONSOB a notice on proprietary-trading platforms in July 2024, and ESMA a public statement on 24 February 2026 (ESMA35-243228190-8024) confirming that novel leveraged products fall inside national product-intervention measures. CONSOB's notice characterised the product as closer to a game than an investment service and recorded complaints that challenges are structured to induce repeat attempts and that agreed profits go unpaid. In the United States, a CFTC consultation on whether evaluation fees constitute commodity-pool participation interests is reported to have opened on 1 August 2026 and to close on 30 November 2026; we cite it as reported, since the consultation document itself postdates most of this paper's drafting. None of this is settled. The operative fact for §13.5 is narrower and is not in dispute: there is no deposit insurance, no compensation scheme, and no supervisory body to which a participant can appeal a refused payout. The counterparty adjudicates, and that is a feature of the current arrangement rather than an allegation about any firm's conduct.

## 10. Scope of the claim

**The claim that no cost-surviving edge exists is false as written, and this paper does not make it.** The research ledger underlying this work contains two candidates that cleared every research gate applied to them before the programme closed, one with its decisive matched control still pending: a leveraged-ETF end-of-day convexity short with an out-of-sample Sharpe ratio of 3.07 over 60 trades, and a holiday-eve de-risking effect with an out-of-sample Sharpe of 3.15 over 53 trades. Both were escalated for deployment review in July 2026. Both survived the pre-specified validation protocol as far as it had been run — C10's decisive matched control was still pending when the programme closed. Under normality and taking the ledger figures at face value, each sits roughly ten standard errors above zero and clears the expected maximum Sharpe over 127 null trials — about 2.6 standard errors — by a wide margin; a full deflated-Sharpe treatment would require the per-trade return series, which the ledger summary does not retain, and heavier tails would narrow the margin. With 53 and 60 out-of-sample trades, that is what can be said.

They are also irrelevant to the evaluation product, and the reason is instructive rather than incidental. Between them they generate approximately nineteen trades per year, worth on the order of $560 annually at one micro contract. They fail the evaluation not on edge quality but on cadence: an evaluation account must resolve within a bounded horizon, and a strategy that trades nineteen times a year cannot reach a $3,000 target inside one. This is the cadence arm of the pincer operating on a strategy that comfortably clears the cost arm.

The defensible claim is therefore: *no edge compatible with the cadence, drawdown, and daily bounds of a retail evaluation account exists in the measured strategy universe.* One sentence of scoping converts a fatal referee objection into a supporting observation, because the existence of gate-cleared edges that fail the evaluation on cadence alone is direct evidence for the pincer rather than against the no-go.

Two further scope statements belong in any submitted version:

- Multi-day and swing horizons are excluded by evaluation rules on the accounts studied, so the no-go is a statement about intraday strategies under those rules and not about the underlying instruments.
- The results concern index futures and their micro variants. The cost law that drives much of the argument — fixed commission being roughly ten times heavier in proportional terms on the micro contract than the full-size contract — is instrument-specific and should not be assumed to transfer.

## 11. Fee sensitivity and the sign of expected value

The account fee is the most visible parameter of the product and the one on which firms compete. This section shows that it is close to irrelevant to the sign of the participant's expected value, which is the paper's second novel contribution.

The term denotes the price of the evaluation account and nothing else. It does not cover per-trade trading friction, which §12.6 shows can flip the sign of a participant's expectation across a range of a few dollars a day. The two are different instruments — the account fee enters the cost term once, trading friction enters the drift on every trade — and only the first is the subject of this section. Writing $\pi$ for the probability of passing an evaluation and $\beta$ for the probability of securing a payout once funded, expected extraction per account is $E = \pi\beta W - \kappa$, where $W$ is the payout size and $\kappa$ the account fee. Both $\pi$ and $\beta$ are barrier-crossing probabilities of the form given in §4.2, each bounded well below one at the participant's measured drift, so their product is compressed into the low single digits.

The theorem follows directly, and its condition should be stated with it: at the participant's measured drift. Setting $\kappa = 0$ leaves $E = \pi\beta W$. With joint-gate probabilities in the low single digits and W on the order of a first payout, expected gross extraction is at most tens of dollars per account; in the direct zero-fee test it is $5.61, which is far below the participant's own transaction and time costs. The sign of E is therefore governed by the gate compression $\pi\beta$, not by $\kappa$. Removing the fee entirely does not change it.

### 11.1 The reward-to-risk frontier

The result has a corollary that is easy to miss, and it runs in both directions. A participant with genuinely zero edge still pays transaction costs on every trade, so the absence of an account fee does not make a zero-drift participant break even; trading friction is a headwind independent of the product's pricing. Note also what the theorem does not claim. It is conditioned on the participant's measured drift. At hypothetical drifts well above anything this program measured, the sign is not invariant to price — Table 12 shows a 44% win rate turning positive on the cheapest account and negative on the most expensive. The claim is that at drifts anyone has actually demonstrated, price is not what makes the expectation negative.

A related invariance concerns the reward-to-risk ratio. The required edge above the driftless baseline is not constant across reward-to-risk: approximately +1.4 percentage points is needed at a ratio of 0.5, against approximately +5 points at 1.5. This is a publishable subsection in its own right and is the correct place to address the referee question of whether the no-go is an artefact of the particular ratio chosen. It is not; the ratio moves the threshold but does not change the sign, for the same reason the fee does not.

## 12. The hedged pair, and what its prohibition reveals

There is an obvious attack on the product, well known among participants and explicitly prohibited by every firm studied. It deserves direct treatment rather than dismissal, because it identifies the correct vulnerability, because the reason it fails is not the reason usually given, and because the prohibition itself is evidence about what the seller believes it is selling.

### 12.1 The construction

Buy two accounts. Take a long position in one and an equal short in the other, in the same instrument at the same time. Whatever the market does, one account gains what the other loses. The winner runs to the profit target and converts to funded; the loser breaches its drawdown and dies, at a cost of one fee. Repeat in the funded stage to reach the withdrawal threshold. The participant expresses no market view at any point.

This is not a separate idea from §5.3 — it is the industrial form of it. Section 5.3 establishes that a capped-downside account carries option value and that pass rates are manufacturable by position sizing at zero skill. The hedged pair is the systematic harvest of exactly that asymmetry: delta-neutral, directionally indifferent, holding the option on both sides at once. If the argument of §5.3 is right, this construction should work. We therefore simulated it.

### 12.2 It is approximately expected-value neutral under symmetric sizing

Mirrored daily profit and loss, both accounts paying costs, both carrying the target and trailing drawdown of the account class studied, ninety-day horizon, 200,000 paths per cell:

| σ/day | cost/day | P(≥1 passes) | P(both) | Cost per funded | Unhedged baseline |
|---|---|---|---|---|---|
| $500 | $20 | 0.422 | 0.000 | $465 | $465 |
| $1,000 | $5 | 0.624 | 0.000 | $314 | $313 |
| $1,000 | $20 | 0.590 | 0.000 | $332 | $331 |

| σ/day | cost/day | P(≥1 passes) | P(both) | Cost per funded | Unhedged baseline |
|---|---|---|---|---|---|
| $1,500 | $20 | 0.672 | 0.000 | $292 | $291 |
| $2,000 | $20 | 0.733 | 0.000 | $267 | $266 |

*Table 22. The symmetric hedged pair against a single unhedged account. Cost per funded account is identical to three significant figures in every cell. Source: hedged-pair simulation, this paper.*

**The hedge doubles the probability of obtaining a funded account and doubles the price paid for it.** Cost per funded account is unchanged in every cell tested, to within Monte Carlo error. Under symmetric sizing and the observed payout ceiling, the construction is a variance transformation, not an edge; §12.4 and §12.5 show the neutrality is not a universal property of it. It converts an uncertain outcome into a more reliable one, which is worth something to a participant constrained by time or by attempts, and worth nothing to one constrained by capital.

### 12.3 Why it is neutral

Two facts do the work, and neither is the one usually cited in practitioner discussion.

**First, the pair is strictly disjoint.** Because equity in the second account is the negative of the first less costs, both accounts reaching +T would require their equities to sum to +2T, when in fact the sum is pinned at minus the accumulated cost. The model implies P(both) = 0 exactly under perfect mirroring, and no joint pass was observed in 200,000 paths in any cell; the table reports the empirical frequency, which is below simulation resolution. It follows that P(at least one passes) is exactly twice the single-account pass probability — bought for exactly twice the fee.

**Second, and less obviously, the winner is not safe.** The profit target exceeds the drawdown allowance, so reaching the target requires a move of $3,000 without a $2,000 retracement from the running peak. A reversal large enough to kill the loser and then turn kills the survivor too. This is why the probability that neither account passes is substantial — between 0.26 and 0.99 across the cells tested, rising sharply as position size falls. The folk description of the hedge assumes one account is guaranteed to run to target. It is not.

### 12.4 Variants

Unequal sizing helps modestly; larger baskets raise the cost per funded account.

| Variant | E[funded accounts] | Cost per funded |
|---|---|---|
| Symmetric pair (1:1) | 0.589 | $333 |
| Asymmetric pair (2:1) | 0.652 | $301 |
| Asymmetric pair (4:1) | 0.715 | $274 |
| Basket, 1 long vs 2 shorts | 0.713 | $412 |
| Basket, 1 long vs 3 shorts | 0.681 | $576 |
| Basket, 1 long vs 5 shorts | 0.371 | $1,583 |

*Table 23. Hedge variants at σ = $1,000/day, cost $20/day. Unequal sizing breaks the disjointness — both accounts can survive — but the gain is modest. Source: hedged-pair simulation, this paper.*

Unequal positions are not perfectly mirrored, so the probability both survive becomes non-zero and the disjointness of §12.3 weakens. At four-to-one the cost per funded account falls from $333 to $274, roughly seventeen percent — real, and an order of magnitude short of what the construction is popularly claimed to deliver. Baskets of three or more are strictly worse: each extra account pays a full fee while contributing a diminishing share of the directional exposure. Two further variants were tested because both run in the participant's favour and either could have overturned the neutrality. Placing the sacrificial leg at the $39 account and the surviving leg at the $98 account yields $419, $327 and $282 per funded account at σ = 500, 1,000 and 2,000 against $465, $332 and $267 for the same-firm pair — neutral, because the cheap firm's $189

activation fee falls due whenever the cheap leg is the one that passes, which is half the time. And a pair aimed at the funded-stage withdrawal threshold rather than the evaluation target returns a probability of at least one payout-eligible account of exactly twice the unhedged figure at every sizing tested: the same linearity as the evaluation stage, two fees for twice the probability.

### 12.5 The payout ceiling is the binding constraint

The hedge fails to produce a business not because it fails to raise the probability of a funded account — it does — but because of what a funded account is worth. Carrying the symmetric pair through the funded gate at the payout rates of Table 3:

| Payout ceiling | No consistency rule | 40% cap | 20% cap |
|---|---|---|---|
| $1,000 (observed) | +$9 | +$9 | −$93 |
| $2,000 | +$215 | +$215 | +$9 |
| $3,000 | +$420 | +$420 | +$112 |
| $5,000 | +$831 | +$831 | +$317 |
| $10,000 | +$1,857 | +$1,857 | +$831 |

*Table 24. Expected value per hedged pair against the payout ceiling. At the observed ceiling the construction is marginal at best and negative under the tightest consistency regime. Source: hedged-pair simulation, this paper.*

**At the ceiling actually offered, the hedged pair earns $9 per $196 committed on the most generous account, and loses $93 on the account with a 20% consistency rule.** Double the ceiling and the same construction earns $215. The payout cap — introduced in §3.5 as an accounting detail separating headline price from true cost — turns out to be the parameter that neutralises the one construction that materially raises the probability of passing the evaluation gate. This is the cost–cadence pincer appearing a third time: the gate can be beaten, and the prize for beating it is set below the cost of doing so.

### 12.6 The limiting case: the cheapest evaluation in the field

A natural objection is that the conclusion is an artefact of the $87 to $228 price range, since §12.5 shows the construction is sensitive to the ratio of cost to prize. We therefore tested the limiting case: Apex Trader Funding, the cheapest and least restricted evaluation we could source, at a promotional price roughly a third of the cheapest account otherwise studied, with rules sourced on 31 August 2026. Its evaluation is the softest in the field — no consistency rule, no minimum trading days, an enlarged contract allowance, and in one variant no daily loss limit. If the no-go were a pricing artefact, it should fail here.

| Daily cost drag (eval / funded) | Eval pass | P(first payout) | Mean payout | EV per account |
|---|---|---|---|---|
| none | 0.413 | 0.1156 | $1,221 | +$28.38 |
| $25 / $10 | 0.380 | 0.0732 | $1,197 | +$3.35 |
| $50 / $20 | 0.348 | 0.0442 | $1,172 | −$11.98 |
| $100 / $40 | 0.288 | 0.0135 | $1,130 | −$25.63 |
| $200 / $80 | 0.184 | 0.0007 | $1,043 | −$29.87 |

*Table 25. Apex 50K, at a $30 evaluation fee, Gaussian daily model. Zero-drift participant, so every figure is pure account option value. Source: two-stage simulation, this paper; firm rules sourced 2026-08-31.*

The Gaussian daily model is coarser than the rest of the paper's simulations. A trade-level rebuild — Bernoulli trades at 1:1.5, six contracts in the evaluation and two in the funded account, $4 per round turn — gives the following, and it is the version the paper relies on.

| Win rate | Risk per trade | Trades/day | Eval pass | P(payout) | Mean payout | EV at $30 |
|---|---|---|---|---|---|---|
| 40.0% | $400 | 4 | 0.378 | 0.065 | $1,207 | −$0 |

| Win rate | Risk per trade | Trades/day | Eval pass | P(payout) | Mean payout | EV at $30 |
|---|---|---|---|---|---|---|
| 40.0% | $250 | 6 | 0.280 | 0.017 | $1,093 | −$25 |
| 42.0% | $400 | 4 | 0.487 | 0.181 | $1,273 | +$82 |
| 42.0% | $250 | 6 | 0.429 | 0.092 | $1,161 | +$16 |
| 44.0% | $400 | 4 | 0.596 | 0.374 | $1,344 | +$269 |

*Table 26. Apex 50K, trade-level rebuild. At the driftless 40% the account is negative at every sizing; it turns positive between 40% and 42%, consistent with the 40.9–41.5% break-even of §7. Source: trade-level two-stage simulation, this paper.*

The trade-level model removes the Gaussian model's cost-free positive cell. At the driftless rate every configuration is negative; the account requires roughly the same edge above 40% that every other account in the paper requires. The cheapest evaluation in the field is not an exception to the threshold, and its softness is, as §12.7 argues, located where it does not matter.

**Cost-free, the account is worth +$28 per evaluation purchased. With realistic friction applied it crosses into negative territory between $25 and $50 of daily cost drag — a handful of round turns per session at the permitted contract size.** This reproduces, at the cheapest price point in the market and with no strategy involved at all, the pattern that ended the first research cycle: 190 surviving configurations without costs and none with them. The cost floor is not a feature of the accounts previously studied. It is the binding constraint at every price observed.

### 12.7 Where the product actually prices itself

The more instructive finding is where Apex places its friction. Every constraint that binds on the hedged construction sits downstream of the evaluation, in the funded account.

1. A permanent safety net. Only profit above the starting balance plus the drawdown allowance may be withdrawn, and the balance must remain above that line after every payout for the life of the account. On the 50K tier this traps the first $2,100 of profit indefinitely, and the minimum request threshold sits $500 above it again.
2. Five qualifying days, each requiring a minimum profit in its own right. Small green days do not count toward the total.
3. A 50% consistency rule applied at payout: no single day may constitute half or more of profit since the last approved payout.
4. A contract allowance that falls to a third of the evaluation allowance on transition to the funded account.

**Items two and three are jointly fatal to the hedged construction.** A delta-neutral pair resolves in a single large move; that is its entire mechanism, and it is why it defeats the evaluation so reliably. The same property produces exactly one substantial profit day, which simultaneously fails the five-qualifying-day count and breaches the 50% consistency test. The reduced contract allowance removes the option of sizing around either. A participant cannot hedge into a payout at this firm under any parameterisation we could construct.

This bears on the puzzle §12.8 takes up. At this firm the prohibition on hedging is not the primary defence and does not need to be, because the payout gate already forecloses the construction mechanically; the prohibition is redundancy on a gate that is already closed.

**The general lesson generalises beyond this firm and is the sharpest single statement of the paper's thesis.** The evaluation is not where the product prices itself. It is the marketed surface, and it is economically permeable — a soft evaluation sells accounts, and a participant who passes one has bought a lottery ticket rather than won anything. The economically binding friction is concentrated in the payout gate, where the participant has already paid, has no remaining leverage, and where the counterparty is also the adjudicator. Any analysis that measures pass rates and stops has measured the marketing.

> *Provenance: Apex's pricing is promotional and, on our reading, recurring rather than one-time; an evaluation that runs beyond its access window incurs the fee again, which our single-payment model does not capture and which moves every figure in Table 25 downward. The funded-stage payout probability is also sharply peaked in position size, falling away on both sides of the maximum reported, so even the cost-free positive cell is a knife-edge rather than a stable region. Both should be verified against the firm's current terms before the figures are relied upon.*

### 12.8 The prohibition, and what it implies

Every firm studied prohibits this construction by contract: hedging across accounts, copy trading between accounts, group or coordinated trading, and account sharing are all excluded, and firms state that they monitor for them and will refuse payment and terminate accounts on detection. The participant has no recourse, since the counterparty is also the adjudicator.

Section 12.7 resolves what would otherwise be a puzzle. Our measurement says the hedged pair is roughly break-even at the observed ceiling, so on our numbers the prohibition appears to protect the firm from very little. The answer is that at the firm with the most explicit payout gating, the construction is already foreclosed mechanically — the qualifying-day count and the consistency test defeat it without any contractual term being invoked. The prohibition is redundancy on a closed gate rather than the primary defence. We note that this reading is available to us only because we sourced a firm whose funded-stage rules are unusually explicit; at firms with looser payout gating the prohibition may well be doing the work directly, and we have not measured that case.

**The prohibition is nonetheless the most direct evidence in this paper for the paper's central characterisation.** A rule excluding delta-neutral positioning is not, on its face, a risk-management rule: a perfectly hedged participant takes no market risk. Whatever the drafters' reasons — which we cannot observe and do not speculate about — the rule's effect is to protect the gate rather than the participant. And a gate that must exclude a strategy expressing no market view is not, by that fact, distinguishing skill. This is an inference about what the clause does, not about why anyone wrote it.

### 12.9 What this simulation does not model

Two omissions remain after the cross-firm and funded-stage variants of §12.4 were measured and found neutral. We assume perfect simultaneous mirrored fills, which runs against the participant rather than for them. And we apply no probability of detection — which, since detection forfeits the payout rather than the fee, would move every figure in Table 24 downward by an amount the firm rather than the participant controls.

## 13. The counterparty model

The analysis to this point has treated the product as a contract executed as written and asked what it is worth to the participant. This section takes the other side. The setting is a principal–agent problem in the sense of Holmström (1979), with one inversion: there a principal designs a contract to elicit unobservable effort, whereas here the participant's effort is largely irrelevant to the seller's payoff and what the participant cannot observe is the contract's own outcome distribution. It asks what the seller's problem is, which instruments the seller controls, and whether the rule geometry documented in §3 is what a seller solving that problem would choose. If it is, the negative expectation established above is consistent with the output of such an optimisation rather than with the difficulty of the market, and the paper's central claim follows.

### 13.1 The seller's problem

The evaluation account is a capped-downside claim: the participant pays a premium, receives a payoff bounded above by the payout ceiling, and loses at most the premium. Its valuation has the structure of the barrier-option problem of Merton (1973), with the trailing drawdown as a knock-out boundary — and §5.3 exploits precisely that structure when it shows the claim can be brought into the money by raising volatility at zero drift.

Write $\pi$ for the probability a purchased evaluation is passed, $\beta$ for the probability that a funded account secures a payout, W for the payout size, and $\kappa$ for the account fee. From §11, participant expectation per account is $E = \pi\beta W - \kappa$. Take the fee stream as the seller's revenue and the payout stream as its only economically relevant liability, ignoring operating, acquisition and processing costs and any ancillary revenue. Per account, the seller's contribution margin under this simplified contract model is

$$\Pi = \kappa - \pi\beta W = -E$$

Under those assumptions the arrangement is exactly zero-sum — a statement about the contract model, not an accounting of any firm's profit and loss. It is worth stating plainly because it disposes of a common framing. The evaluation is sometimes described as a talent-identification mechanism whose revenue funds a search for traders to back with real capital. Under that description the seller would be indifferent to $\Pi$ and would set the gate to maximise the information content of passing. The observed gate does not appear to do so — §5.4 finds it a weak and confounded signal, and §9.1 reports that 0.71% of funded participants reach a live account — and §13.3 sets out what the model says it would maximise instead.

The seller's instruments are the profit target T, the drawdown allowance d, the consistency cap c, the minimum-days requirement m, the payout ceiling W, and the price $\kappa$. The first four enter only through $\pi$ and $\beta$, via the first-passage factors of §4.2. The seller therefore does not choose $\pi$ and $\beta$ directly; it chooses barrier geometry, and the probabilities follow.

### 13.2 The participation constraint is written in beliefs

A seller cannot simply set $\Pi$ as large as it likes, because the participant must be willing to buy. If participants were informed and risk-neutral they would purchase only when $E \geq 0$, which is exactly $\Pi \leq 0$. On that constraint the business does not exist. It follows that the business exists only where participants' beliefs about the gate differ from the gate.

Let $\hat{\pi}$ and $\hat{\beta}$ denote the participant's subjective probabilities at the point of purchase. Participation requires $\hat{\pi}\hat{\beta}W \geq \kappa$. Substituting into $\Pi$ gives the following.

**Proposition 5 (the belief gap bounds the take).** Under voluntary participation and the simplified contract model, seller contribution margin per account satisfies

$$\Pi \leq W \cdot ( \hat{\pi}\hat{\beta} - \pi\beta )$$

Proof. Participation gives $\kappa \leq \hat{\pi}\hat{\beta}W$. Substituting into $\Pi = \kappa - \pi\beta W$ yields $\Pi \leq \hat{\pi}\hat{\beta}W - \pi\beta W = W(\hat{\pi}\hat{\beta} - \pi\beta)$. □

**The seller's objective is therefore not to minimise the joint gate but to maximise the gap between the perceived gate and the actual one, scaled by the payout.** This reframes every design question in the paper. A gate that is visibly hard suppresses $\hat{\pi}\hat{\beta}$ along with $\pi\beta$ and earns nothing. A gate that is genuinely easy earns nothing either. The profitable configuration is one that is easy where the participant can see it and hard where the participant cannot.

### 13.3 Which factor to give away

The two factors are not symmetric in what the participant can learn about them, and that asymmetry determines the design.

The pass probability $\pi$ is cheaply observable. A participant learns it by purchasing an evaluation and finding out, and communities of participants pool the observation. Belief $\hat{\pi}$ therefore converges on $\pi$ quickly, and the term $\hat{\pi} - \pi$ is not a durable source of profit. The payout probability $\beta$ is not observable on the same terms. It is realised only after the fee is sunk, only by the subset who passed, over a horizon of weeks, and conditional on a compound of requirements — a safety net, a qualifying-day count, a consistency test, a ceiling — whose joint effect is not legible from any one of them. The model allows $\hat{\beta}$ to be anchored instead on the experience of the evaluation, which was easy; whether it is so anchored is the unmeasured quantity of §13.6.

**If the seller can influence how observable each factor is, the model implies an incentive to give away $\pi$ and gate on $\beta$.** A soft evaluation is not a concession; it is the product's advertising, and it is nearly free to provide because $\beta$ gates the payout regardless. This is consistent with the design the measurements find. §12.6 documents a firm whose evaluation is the softest in the field — no consistency rule, no minimum days, an enlarged contract allowance, passable in a single session — and every binding constraint located downstream in the funded account. §5.3 supplies the mechanism from the participant's side: a zero-skill participant manufactures a pass probability near 0.40 by sizing alone, so passing is both common and uninformative. What §12.7 observes is consistent with the model.

The frequency asymmetry of §6 admits the same reading. The evaluation rewards a fast, lumpy cadence and the funded account punishes it, by a factor of nine in the joint gate. A participant who learns to pass has learned a behaviour that is counterproductive at the stage that matters, and has learned it at the seller's expense of nothing.

### 13.4 Rule or price

The seller can close an avenue by pricing it or prohibiting it. The choice is informative because prohibition is costly to monitor, invites dispute, and signals the avenue was worth closing.

The hedged pair of §12 is the natural test case, since it is the one construction that reliably defeats the evaluation gate. A delta-neutral participant sets $\pi$ to nearly one at the cost of a second fee. In the language of Proposition 5 this collapses the $\pi$ side of the belief gap: there is nothing left to be wrong about. The seller's response is available in both forms, and both are observed. The payout ceiling prices it — §12.5 shows the construction earning \$9 per \$196 committed at the observed ceiling and \$215 at twice that ceiling, so the cap alone is nearly sufficient. The prohibition then closes it by rule. At the firm of §12.6 the qualifying-day count and consistency test foreclose it mechanically before the prohibition is invoked at all.

**That three distinct mechanisms close the same avenue is evidence consistent with a closely aligned composite gate, though it does not establish optimisation.** A ceiling at which a delta-neutral participant nets nine dollars, a payout gate that forecloses the same construction mechanically, and an explicit contractual exclusion of it are three independent closures of one avenue. We make no claim about how they came to coincide; firms converge on terms through competition, imitation and experience as readily as through calculation, and the historical record is not available to us. The observation is that the resulting contract behaves consistently with optimisation of the seller's contribution margin, and that this is the closest the paper comes to agreement with the seller about the mechanism.

### 13.5 Discretionary non-payment

One instrument remains outside the barrier geometry. The seller adjudicates its own payouts, and payment may be refused or delayed on grounds it defines and evaluates. Introducing $\delta$ as the probability that an earned payout is not received, participant expectation becomes $E = \pi\beta(1 - \delta)W - \kappa$ and the seller's take rises correspondingly.

This is not an inference. The firm of §12.6 states in its own published risk disclosure that reward payouts are discretionary and subject to eligibility and compliance, and that all activity occurs in a simulated environment using virtual funds, with no real capital at risk and no live market execution. Both statements are the seller's, not ours, and both are load-bearing here: the first establishes $\delta$ as an instrument the seller reserves explicitly, and the second forecloses the talent-identification framing dismissed in §13.1 — there is no live capital at the end of the pipeline to identify traders for.

A second instance is contractual rather than discretionary. At Bulenox, under terms retrieved on 31 August 2026, declining the transition from the intermediate funded stage to the final one closes the account with no payout and forfeits any unpaid profit balance. The participant's accrued earnings are contingent on accepting a subsequent contract whose terms are set by the counterparty. That is not non-payment at discretion; it is non-payment by design, and it belongs in $\delta$ alongside the discretionary case.

The term deserves separate treatment because it is unlike the others in three respects. It carries no visible price, so it does not enter $\hat{\pi}\hat{\beta}$ and widens the belief gap directly. It is exercised after the participant has performed, when there is no remaining leverage. And it is adjudicated by the counterparty, so the participant's recourse is to the party declining to pay.

Instances of the instrument in use are documented rather than inferred. Finance Magnates (2024) records one of the largest futures firms falling months behind on payouts and, at one point, having processed roughly thirty percent of outstanding trader payments; the same firm subsequently conditioned payout on the trader submitting two full trading days of screen and camera recording. A second firm marketing itself on a record of zero denied payouts is recorded refusing one. And the cross-firm dataset of §9.2 reports that fifty-five percent of participants who reached a funded account never received a payout from it — a figure that combines participants who failed the funded gate with those whose payouts were declined, and which the data cannot separate. Whatever $\delta$ is, it is not zero, and even a modest value is material against a break-even threshold sitting under a point above the driftless baseline.

The same instrument closes the residual argument left open by §5.3. If the geometry permits a pass rate near 0.40 with no skill, why is variance-farming not itself a business? Because the behaviour is detectable and the seller declines to pay for it. No model of barrier geometry captures a terms-of-service enforcement action, and the participant who plans around the geometry has planned around the wrong instrument.

### 13.6 What the model does not settle

Proposition 5 bounds the seller's take by the belief gap but does not derive the gap, which is an empirical quantity we have not measured. Doing so would require eliciting $\hat{\pi}$ and $\hat{\beta}$ from participants at the point of purchase, and we have no such data. The proposition is therefore a constraint on any explanation of the industry's economics rather than a measurement of them, and it should be read as identifying what would have to be true rather than as establishing that it is.

Three further limits should be stated. We treat the seller as a single agent maximising per-account profit, where a real firm faces reputational constraints, competition on visible terms, and the possibility of regulatory attention, all of which bound $\delta$ from above in ways we have not modelled. We take $\delta$ from a single self-reported source. And we have not established that any particular firm reasons in these terms; the claim is that the observed rule geometry is what such reasoning would produce, which is weaker than a claim about intent and is all the evidence supports.

**What the model does settle is the question §13.1 opened with.** A seller indifferent to $\Pi$ and interested in identifying traders would set a gate that is informative about skill. The observed gate is a weak and confounded skill signal at the evaluation stage, and its binding constraints sit where the participant cannot price them. Those are the choices the model predicts for a seller maximising $\Pi$ subject to participation, and the negative expectation documented throughout this paper is their arithmetic consequence rather than a market outcome; whether any firm arrived at them by that reasoning is not something the evidence can say.

## 14. Conclusion

The retail proprietary-trading evaluation is a contract whose geometry rewards different behaviour at each of its two stages and whose evaluation stage can be passed without skill. Under end-of-day trailing, the evaluation is best attacked with a fast, lumpy cadence and the funded account with a slow, steady one, by a factor of nine in the joint gate; a participant who learns to pass has learned a behaviour that is counterproductive where it matters. Position sizing alone yields a pass probability near 0.40 at zero edge against a measured cohort rate of 0.168, so passing is not, by itself, reliable evidence of skill, and any economically consequential selection not explained by the pass rate occurs at the payout gate, where the participant has already paid and cannot price what remains. The sector's disclosure pattern — pass rates published far more often than payout rates — and the convergence of inverted rule architectures on comparable outcomes are both consistent with a seller whose contribution margin is bounded by the gap between what the participant believes about the gate and what the gate is.

The same geometry produces negative expected value within the strategy universe we measured. Across the three accounts studied in detail and two implementations, break-even lies between a 40.5% and 41.5% win rate at 1:1.5 net of costs against a driftless 40.0%; the threshold rises to roughly 44% on 20%-consistency accounts and above 46% on high-fee accounts. Two constraints put that out of reach for every candidate we tested: a transaction-cost floor exceeding every gross per-trade edge surviving the protocol, and a joint gate the account geometry compresses to at or below the seller's own break-even line. Neither is a parameter the participant controls.

Three of the paper's findings extend beyond the immediate question. The first is that evaluation pass rates are manufacturable by position sizing alone, which means the evaluation pass rate cannot function as a standalone skill filter and any consequential selection not explained by it must occur downstream at the payout gate. The second is that the two stages demand opposite trade cadences by a factor of nine in the joint gate, which supplies a mechanism for an industry practice that practitioners follow and do not explain. The third is that at the drifts anyone has demonstrated, variation in the account fee across its observed range does not overturn the negative sign, which relocates the analysis from pricing, where the industry competes, to geometry, where it does not.

We close on the scope condition, because it is the part most likely to be misread. This paper does not claim that systematic edges are unavailable to small participants. Two candidate edges in the underlying ledger cleared every gate that had been run and post out-of-sample Sharpe ratios above three. They fail the evaluation product on cadence, not on quality — nineteen trades a year cannot clear a $3,000 target within a bounded evaluation horizon. That two out-of-sample-positive candidates are excluded by the product's structure while a zero-skill participant reaches a pass rate near forty percent is, in the end, the most compact statement of what the product selects for.

# Appendix A. Verification protocol

All simulation code, the figure-generation scripts, and the build script producing this document are public at https://github.com/nicholasbhall/gate-design-prop-evals. The checks below are what a reader with that repository can run.

## A.1 Reproduction checks

- Re-run the phase-diagram script unchanged and confirm Tables 5–8 reproduce within Monte Carlo error; exact reproduction is expected on the same numpy version and seeds; an independent implementation should agree within Monte Carlo error.
- Cross-check the cell ($\mu = 5$, $\sigma = 90$, $f = 5$) against an independent Monte Carlo in the backtest engine.
- Confirm the analytic anchor: the two-barrier fixed-level probability at $(\mu, \sigma, f) = (5, 100, 10)$ should equal 0.871 under the expression stated in §4.3, and the driftless limit should return $d/(d+T) = 0.400$.
- Re-run the adversarial-test script with fixed seeds and confirm the out-of-sample win rate of 38.54% and mean of −$32.14 per trade.

## A.2 Standing methodological commitments

- Selection and tuning on in-sample data only; one out-of-sample validation per candidate; never re-select after seeing out-of-sample results.
- A zero-edge control through the identical pipeline is mandatory. No result counts without it.
- Edges, pass rates, and Sharpe ratios are outputs, never targets. A stated target win rate has no honest execution.
- Null results are valid outputs and are reported.
- Firm parameters are sourced from firm sites and carry retrieval dates; where a required rule cannot be sourced it is flagged rather than assumed.

## Appendix B. Outstanding work

Every reference in this paper was verified individually against the published record; none is generated, inferred from a secondary mention, or reconstructed from memory.

### B.1 Citation status

No claim in this paper rests on an unverified source. Every scholarly reference was checked individually against the published record; every firm parameter and statistic carries a retrieval date; every trade-press figure is identified as such in the text. Two figures that earlier drafts carried on practitioner self-report — a confiscation rate and a sector firm count — were removed and replaced with documented instances or with claims the paper does not need.

### B.2 Future work

Of the items recorded in earlier drafts, five have been done and folded into the body: the skill gradient (§5.4), the distributional robustness checks (§5.6), an indicative deflated-Sharpe treatment of the near-survivors (§10), the cross-firm and funded-stage hedge variants (§12.4), and the trade-level rebuild of the Apex model (§12.6). What remains is work the paper cannot do with the data it has.

**Measurement of the belief gap.** Proposition 5 bounds the seller's contribution margin by the gap between perceived and actual gate probabilities, and §13.6 concedes the gap is unmeasured. Eliciting participants' subjective pass and payout probabilities at the point of purchase, and comparing them with the published figures of §9, would convert the proposition from a constraint on explanations into a measurement. It is also the test that would distinguish the shrouding account from the alternatives listed in §12.8. It requires a survey, which this paper does not conduct.

**A full deflated-Sharpe treatment of the two near-survivors.** This is supplemental rather than load-bearing. The indicative calculation in §10 is what the paper relies on, and §10's scope claim does not depend on either candidate being genuine: both are excluded from the evaluation product on cadence, and if a full treatment showed either to be spurious the concession in §10 would weaken while the no-go would not. A proper treatment needs skew, kurtosis and autocorrelation of the per-trade returns, which the archive holds and the ledger summary used here does not.

**Seller-side cost accounting.** The seller model of §13 treats payouts as the only liability. A version carrying customer acquisition, payment processing, platform and operating costs would bound the contribution margin from below as well as above, and would let the paper say something about the sector's economics rather than only about the contract's. The inputs are not public.

**Retry correlation.** Evaluations are modelled as independent draws, whereas a fixed strategy in a persistent regime produces correlated outcomes across attempts. This raises the variance of every account-level figure without changing its mean, and a participant buying several accounts should expect wider dispersion than the independent model implies. Stating the magnitude requires a regime model calibrated to the instrument, which §5.6's volatility-switching variant approximates but does not resolve.

**Data the paper cannot generate.** A second futures firm publishing full cohort statistics on the same denominators as Topstep would do more for §9 than any further simulation, and the CFTC consultation closing in November 2026 may produce disclosure the sector has so far withheld.

## Appendix C. Cycle II geometry sweep

Ten mechanical geometries, gross edge in index points per trade and net dollars per trade per contract after the mandated $3.00 round-turn cost. All ten are net-negative; eight are gross-negative. This table is the cost-floor gate of §2.3 applied to a complete sweep, and no firm parameter enters it.

| Geometry | Gross pts/trade | Net $/trade/contract |
|---|---|---|
| VWAP mean reversion | +0.318 | −2.37 |
| 5-minute gap breakout | +0.317 | −2.37 |
| 5-minute 10/10 breakout | −0.174 | −3.35 |
| 15-minute 20/20 breakout | −0.184 | −3.37 |
| Prior-range fade | −0.190 | −3.38 |
| Opening-range fade | −0.195 | −3.39 |
| VWAP reversion at R:R 2.0 | −0.376 | −3.75 |
| 5-minute run breakout | −0.482 | −3.96 |
| VWAP reversion at R:R 1.5 | −0.507 | −4.01 |
| 30-minute 15/15 breakout | −0.716 | −4.43 |

*Table 27. The complete Cycle II geometry sweep. Best net result is −$2.37 per trade per contract. Source: research ledger, Cycle II Track B.*

## References

Each reference below was verified individually against the published record.